%% file: 36113.tex
\documentclass{aa} 

\usepackage{epsfig,natbib,url,twoopt}
\usepackage{graphicx}
\usepackage[varg]{txfonts}
\usepackage{hyperref}          
\usepackage[usenames]{color}
\usepackage{acronym}      
\hypersetup{
  colorlinks=true,  
  urlcolor=blue,    
  linkcolor=red     
}

\def\online#1#2{\href{http://www.staff.science.uu.nl/~rutte101/rrweb/rjr-pubstuff/rbecontrails/#1}{#2}}

\makeatletter
\def\aa@manuscriptname{%
 \hfill Volume 632~~~ Article number A96 \hfill
  \href{https://doi.org/10.1051/0004-6361/201936113}
       {doi.org/10.1051/0004-6361/201936113}}
\makeatother

\input{rrmacros-pub}

\def\revapar{}                 
\long\def\reva#1{#1}           
\def\revbpar{}                 
\long\def\revb#1{#1}           
\long\def\revc#1{#1}           

\def\rmit#1{#1}                 
\def\specchar#1{{\textsc{#1}}}  

\def\PubA{\href{http://ui.adsabs.harvard.edu/2009ApJ...705..272R/abstract}
  {Pub\,A}}
\def\linkpubapage#1#2{\linkadspage{2009ApJ...705..272R}{#1}{#2}}
\bibnote{2009ApJ...705..272R}{(Pub\,A)} 
\def\PubB{\href{http://ui.adsabs.harvard.edu/abs/2017A&A...597A.138R/abstract}{Pub\,B}}
\def\linkpubbpage#1#2{\linkadspage{2017A&A...597A.138R}{#1}{#2}}
\bibnote{2017A&A...597A.138R}{(Pub\,B)} 

\begin{document}

\title{Solar H-alpha features with hot onsets}
\subtitle{IV. Network fibrils}

\author{Robert J. Rutten\inst{1,2,3} 
\and 
Luc H. M. Rouppe van der Voort\inst{2,3}
\and 
Bart De Pontieu\inst{4,2,3}
} 

\institute{\LA \and \ITA \and \RoCS \and \LMSAL}  

\date{Received \revb{16 June 2019}     
  /  Accepted \revb{19 August 2019}
  /  Published 17 December 2019 }

\abstract{ Even in quiet areas underneath coronal holes the solar
chromosphere contains ubiquitous heating events. 
They tend to be small scale and short lived, hence difficult to
identify. 
Here we do not address their much-debated contribution to
outer-atmosphere heating, but their aftermaths.
We performed a statistical analysis of high-resolution observations in
the Balmer H\hspace{0.1ex}$\alpha$ line to suggest that many slender
dark H\hspace{0.1ex}$\alpha$ fibrils spreading out from network
represent cooling gas that outlines tracks of preceding rapid type II
spicule events or smaller similar but as yet unresolved heating agents
in which the main gas constituent, hydrogen, ionizes at least
partially.
Subsequent recombination then causes dark H\hspace{0.1ex}$\alpha$
fibrils enhanced by nonequilibrium overopacity.
We suggest that the extraordinary fibrilar appearance of the
H\hspace{0.1ex}$\alpha$ chromosphere around network results from
intermittent, frequent small-scale prior heating.
}


\maketitle

\section{Introduction}\label{sec:introduction}
The solar chromosphere is finely structured.
\citetads{1868RSPS...17..131L} gave it its name for the colorful
appearance of the prominence lines he saw rimming the entire Sun
off-limb, primarily \HI\ Balmer and \HeI\ lines (he also detected and
named helium).
\revc{Solar imaging in Balmer \Halpha\ }shows that
\revc{the on-disk counterpart of} Lockyer's chromosphere consists of
dense canopies of slender fibrils \revc{where} there is some magnetic
activity (\eg\ \citeads{1974soch.book.....B}) and that these fibrils
are highly variable with time.  
Thus, the solar chromosphere is a dynamically fibril-structured
envelope around the thinner photospheric shell. 
In the meantime it became clear that it is the domain where hydrogen
ionizes and where coronal heating and mass loading are rooted. 

\reva{So-called type II spicules are an important chromospheric ingredient
in ``quiet'' areas outside active regions.
Since their first identification by
\citetads{2007PASJ...59S.655D},
they have attracted much attention and debate regarding
their nature, their physical drivers, and their role as quiet-Sun
sources of mass and energy loading of the corona.
We refer to the recent studies of
\citetads{2017ApJ...845L..18D}, 
\citetads{2018ApJ...856...44A}, 
\citetads{2018ApJ...860..116M},
and \citetads{2018ApJ...857...73C} 
for a current overview and references, but here we do not discuss these
aspects further because we instead address another topic: their
chromospheric aftermaths.}  

The analysis presented here targets \Halpha\ fibrils around moderately
active network bordering coronal holes.
It is a sequel to \citetads{2009ApJ...705..272R} 
(henceforth Pub\,A) and
\citeads{2017A&A...597A.138R} 
(henceforth Pub\,B) and uses the same \Halpha\ imaging sequences, both
obtained with the \acl{CRISP} (CRISP,
\citeads{2008ApJ...689L..69S}) 
at the \acl{SST} (SST,
\citeads{2003SPIE.4853..341S}). 
The first study established that ``\acl{RBE}'' (RBE) fibrils in the
outer blue wing of \Halpha\ are on-disk manifestations of off-limb
type II spicules similar to those that were found before in \CaIR\ by
\citetads{2008ApJ...679L.167L}. 
The same was then found for ``\acl{RRE}'' (RRE) fibrils in the outer
red wing of \Halpha\ by
\citetads{2013ApJ...769...44S}. 
The introduction to \PubB\ reviewed these and also magnetoacoustic
field-guided ``dynamic fibrils'' emanating fairly upright from plage
and network, similar but shorter ones in sunspot chromospheres, and
``long fibrils'' , which appear to span from magnetic roots in network
or plage far out over adjacent internetwork cells. 
We refer to this summary rather than repeating it here.

Very thin long fibrils called ``\acl{SCF}s'' observed in the cores of
the \CaII\ \HK\ lines were studied by
\citetads{2017ApJS..229...11J} 
after being reported earlier by
\citetads{2009A&A...502..647P}. 
These may also be on-disk representations of the type II spicule 
phenomenon, as are the comparable long, thin \CaIIH\ ``straws'' close
to the limb described by
\citetads{2006ASPC..354..276R},  
but this has not yet been verified.

\revapar

\reva{This study of spicule II aftermaths} is inspired by the
example case in \PubB\ of a striking long, thick, dark \Halpha\ fibril
that appeared minutes after a hot disturbance was launched from a
small patch of moderately active network \reva{that} was called a
``\acl{PHE}'' (PHE).
It appeared similar to an RBE marking the launch of a type II spicule
but extended unusually far.
It was recognized in the far blue \Halpha\ wing from RBE-like
combination of large blueshift and large broadening and became visible
also in 304, 171, 193, and 94\,\AA\ images
from the \acl{AIA} (AIA) on board the \acl{SDO} (SDO) as an
accelerating PHE, an extending thin long jet-like feature that reached
high temperatures, \reva{implying hydrogen ionization} along its trajectory. 
Subsequently, minutes later, a dark \Halpha\ fibril appeared that outlined
the PHE trajectory rather like the contrails marking preceding
passages of airplanes on our sky. 
The dark fibril then retracted back to the original launch site with
a conspicuous \Halpha\ core redshift.
This \PubB\ case exemplified a heating event reaching \revapar
hydrogen ionization 
along its track, followed by a return aftermath of back-flowing
cooling gas with hydrogen recombination.  

Naturally, the \PubB\ case led to the suspicion that similar but less
fierce type II spicule launches manifested by RBEs and RREs may be
important contributors to the fibrilar scenes that are so emphatically
present in \Halpha\ images, in this case, long fibrils around network. 
\PubB\ reported searches for more such occurrences. 
A few more cases were indeed detected and reported, also at nearly the
same location which suggests recurrence, but the conclusion was that the
fibrilar \Halpha\ scenes are too dynamic, too small scale, and too
confused for easy one-to-one identification of line-core fibrils
appearing a few minutes after blue-wing RBEs.
The impression was that the slenderest fibrilar features showed best
correspondence, but the issue was postponed to future analysis.  
We take it up here.

Thus, our questions are whether the appearance of contrail fibrils and
return aftermath fibrils after similar PHE launches are common, and in
particular, whether RBEs and RREs represent such launches. 
Regardless of the issue whether these chromospheric phenomena
contribute significant heating and mass loading to the outer
atmosphere, they do represent PHEs because they extend accelerating away
from network at increasing temperature (\eg\
\linkadspage{2013ApJ...764..164S}{3}{Fig.~1} of
\citeads{2013ApJ...764..164S}), 
with corresponding type II spicule tips often hotter than 20\,000~K
and sometimes reaching million-degree temperatures
(\citeads{2011Sci...331...55D}, 
see also
\citeads{2016ApJ...820..124H}). 

If they indeed are followed by dark \Halpha\ contrail fibrils, then the
latter constitute a category not present in the numerical \Halpha\
studies of Leenaarts \etal\
(\citeyearads{2012ApJ...749..136L}, 
\citeyearads{2015ApJ...802..136L}), 
whose Bifrost simulation
(\citeads{2016A&A...585A...4C}) 
does contain \Halpha\ fibrils connecting two opposite-polarity network
patches but few fibrils emanating away from the network, and indeed
does not contain RBEs either.  
Because the latter are ubiquitous around network, with only half-minute
lifetimes but \reva{a recurrency} rate of 0.7/min per location where they
occur (\citeads{2013ApJ...764..164S}), 
RBE- and RRE-produced fibrils may contribute much to the fibrilar
appearance of the \Halpha\ chromosphere around network and represent a
direct consequence of and companion to small-scale dynamic heating in
the chromosphere outside active regions.
Our results below suggest that this is the case.

We use the same data as in \PubB,\ but proceed from its single-case
study to full-field and full-duration analysis.
In addition, we use the \PubA\ \Halpha\ data with which
\citetads{2009ApJ...705..272R} 
established \Halpha\ RBEs as type II spicule manifestations.
Both data sets are suited because they combine high resolution
with good wavelength coverage in \Halpha\ at high cadence while
targeting isolated network patches near disk center
that fit within the small SST field of view. 

The next section summarizes the observations and \reva{our} tools.
Section~\ref{sec:results} presents and describes figures that
represent the core of this study. 
We interpret them in Sect.~\ref{sec:interpretation} and then add
discussion and our conclusion.


\begin{figure*}
  \centering  
  \includegraphics[width=\textwidth]{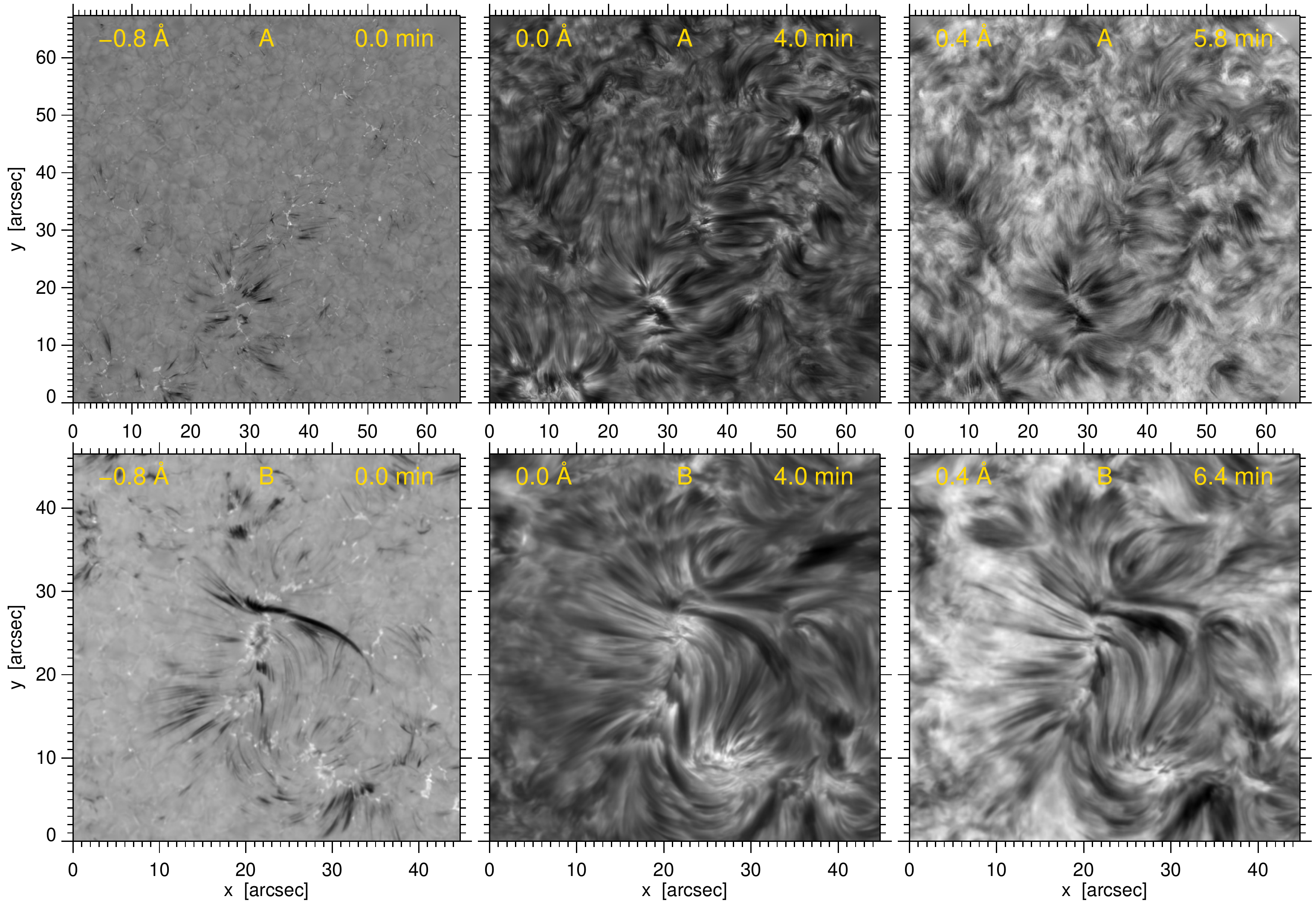}
  \caption[]{\label{fig:overview} Full-field overviews \reva{for
  data~A ({\em upper row\/}) and data~B ({\em lower row\/})}.
  \revapar {\em First column\/}: \Halpha\ blue-wing intensity at
  $\Delta \lambda \tis -0.8$~\AA. 
  {\em Second column\/}: \Halpha\ line-center intensity 4~min later.
  {\em Third column\/}: \Halpha\ red-core intensity at
  $\Delta \lambda \tis +0.4$~\AA\ \reva{about} 6~min after the first
  column.
  Each panel is bytescaled independently.
  The data~A images start at the best-seeing moment.
  The data~B images sample the \PubB\ PHE, contrail fiber, and return
  aftermath. 
  The timings are marked by triangles in Fig.~\ref{fig:seeing}.
  \reva{Online blinkers for these images \revbpar (for page flipping
  per pdf viewer in single-page,} \revb{fit-to-page, or full-screen
  mode): \online{figure01_blink12.pdf}{1-2},
  \online{figure01_blink13.pdf}{1-3},
  \online{figure01_blink45.pdf}{4-5},
  \online{figure01_blink46.pdf}{4-6})}.
  }
\end{figure*}

\section{Observations and methods} \label{sec:observations}

\paragraph{Data A and B.}
We used the \Halpha\ spectral imaging sequences of \PubA\ and \PubB\
and call them data~A and data~B, respectively.
Both were collected with the CRISP at the SST, and both targeted active
network in a very quiet area near disk center that bordered a coronal
hole (Fig.~\ref{fig:overview}).   
In data~B it contained a small pore that was present hours before but
nearly vanished meanwhile.
The observing durations were 24 and 72~min at spectral-scan cadences
6.7\,s and 11.5\,s.
We refer to \PubA\ and \PubB\ for further detail on these observations
and their reduction. 
The other data reported there are not used here.

In \PubA\ an automated detection procedure was used to identify 608
RBE features in the 212 spectral image scans of the data~A \Halpha\
sequence.
Each detection represented a single-image snapshot part of an RBE,
including multiple detections of the same RBE because their lifetimes
were often longer than the 6.7\,s imaging cadence (up to a minute,
histogram in \linkpubapage{10}{Fig.~13} of \PubA).
For each detection the skeleton of the RBE feature was determined and
stored; this skeleton list is also used as input here.

\paragraph{Strous-format scatter correlation.}
In \reva{many figures we show} pixel-by-pixel scatter \reva{relations
between two image quantities} in an informative format based on
Figs.~13.2--13.3 of
\citetads{1994PhDT.......347S} 
\reva{and named after him.
It is here} obtained with IDL programs by A.G.\ and H.W.~de Wijn
\reva{(available under IDL/cubelib at the \wlRRtop{first author's
website})}. 
In this format, sample density contours are shown instead of individual
symbols per pixel-pair sample \reva{for two reasons.
The first is that when plotting} many samples, these overlap and
saturate into non-informative solid black. 
Contours \reva{then} represent a better rendering of \reva{their}
joint probability density.
\reva{The second reason is that plotting pixel-pair samples with a
linear fit quantifying the Pearson correlation coefficient defined for
linear dependence in the presence of noise is less applicable to
typical solar scenes in which multiple and diverse agents cause
varying responses, including nonlinearities.
Plotting sample contours then permits recognizing characteristic
behavior of particular agents and features, also rare ones, rather like
interpreting mountain relief using elevation contours on a
topographical map from which we glean some terms.

The first such figure (Fig.~\ref{fig:strousdemo}) shows the format for
easy-to-interpret agents and serves as introduction to it.
} The outer contour is set at a density at which individual samples
start to merge, as shown by samples outside it. 
\reva{The curves along the right-hand and top sides show the
normalized sample-density distribution per axis quantity.
The summit of the contour mountain, marked with a cross, represents
the most common value pair.
Its location may differ from the maxima of the distributions because
these pixels do not necessarily combine.}   

The two dashed curves crossing near the \reva{summit} specify the
first moments per axis bin, for cuts along columns and rows,
respectively.
They are Cartesian at no correlation (with contours in an elliptical
\reva{or circular} bull's-eye pattern) but come together toward
\reva{slanted lines} at high correlation (to the lower left and upper
right) or anticorrelation (to the lower right and the upper left). 
\reva{When the two quantities share the same scales, these 1:1
correlation lines lie along the diagonals.}

\reva{The labels in the top left corner specify the data set and the
time lag between the samplings of the two quantities. 
The numbers in \revb{the} top right corner specify the total number of
pixel pairs and the density increase factor between successive
contours.
The numbers at the bottom specify linear Pearson correlation
coefficients, at the left for all sample pairs, at the right for the subsets
constituting the four quadrants of the diagram around the mountain
summit. 
At high linear correlation or anticorrelation, the latter are close to
the overall coefficient along the corresponding diagonal direction; they are
small in the other direction.
These partial quadrant coefficients often furnish better
quantification for different agents than the overall coefficient, but
likewise they are strongly weighted toward the high densities near
the summit.
The degree of association at low densities is indicated best by the
angle between the outer parts of the moment curves, but rare features
may affect outer contours only locally.

In various figures including the last panel of
Fig.~\ref{fig:strousdemo}, we apply the format not to single-image
comparisons (instantaneous or time-delay pairs) but to the whole data
sequences or their best-seeing subsets.
Each successive image pair then represents an additional set of pixel
comparisons, which increases the statistics.
Long-lived or often-occurring pair combinations earn higher density in
this temporal summing.}

\paragraph{Showex browser.}
Our main activity in this study consisted of detailed inspection of
the \Halpha\ data searching for line-center contrail fibrils and
return aftermaths after RBEs by blinking enlarged parts of the
spectral \Halpha\ images while varying the sample wavelengths and time
separations. 
In this case, we did not use the CRISPEX browser of
\citetads{2012ApJ...750...22V}, 
but a new one called SHOWEX that was developed by the first author and
is available on \wlRRtop{his website}. 
It is less versatile than CRISPEX in not showing spectral time charts
and offering fewer measurement tools, but it permits fast blinking
with sliders modifying sampling wavelengths and time delays, offers
easy zoom-in to detail, \reva{permits temporal averaging,} and accepts
many different files of different types.
It also shows instantaneous scatter plots in the Strous format, with
\reva{live marking of the scatter-plot location for the pixel pair
under the cursor in one of the images.}
Figure~\ref{fig:selecttracks}, \revapar its online movie versions,
\reva{and the online image blinkers for many figures}
represent shorthand emulations of our extensive inspections.

\begin{figure*}
  \centering
  \includegraphics[width=0.98\textwidth]{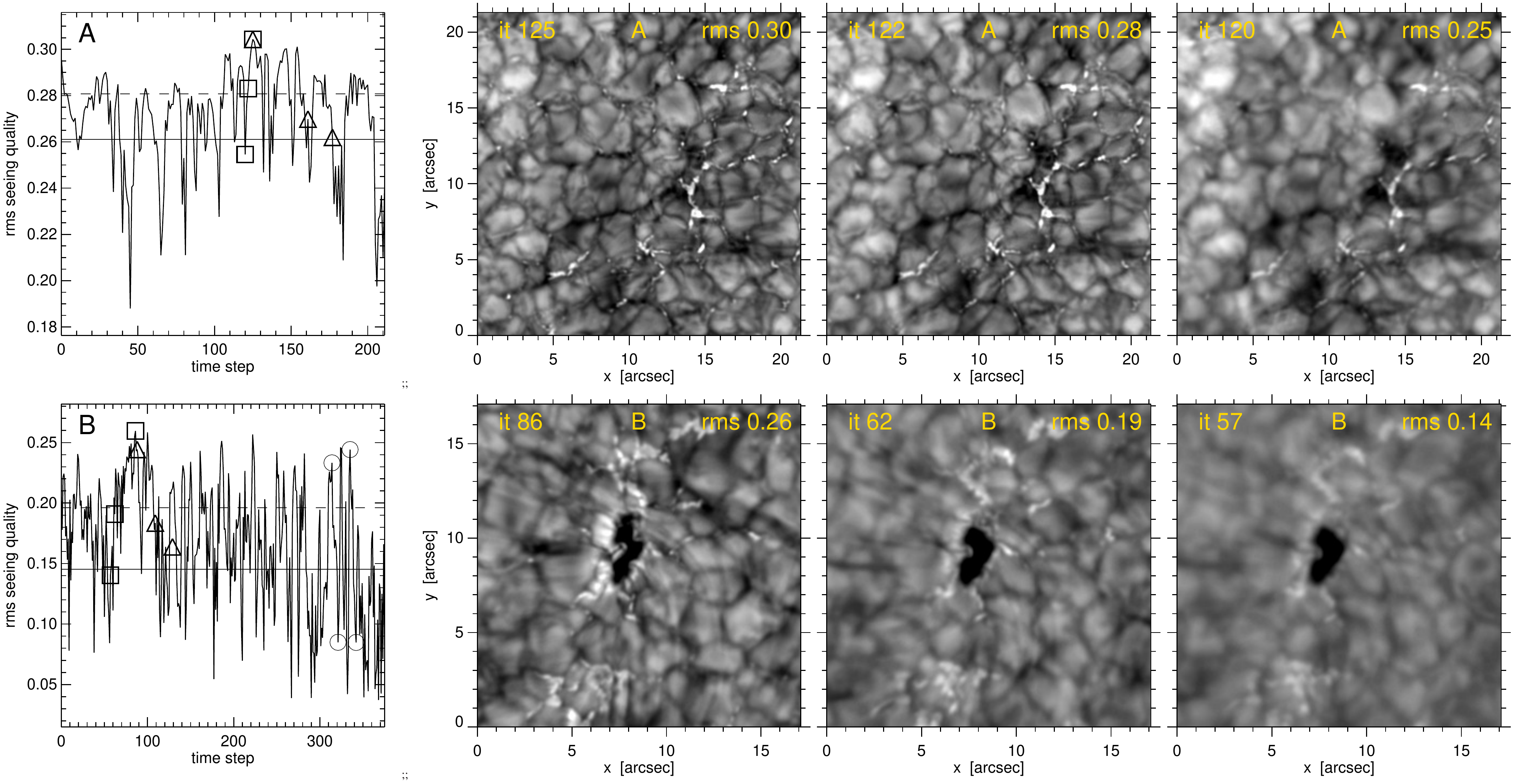}
  \caption[]{\label{fig:seeing} 
  Seeing quality \reva{for data~A 
  ({\em upper row\/}) and data~B
  ({\em lower row\/}).} \revapar 
  {\em First column:\/} root mean square contrast variation 
  \revapar against time for the wide-band \Halpha\ image
  sequences. 
  \reva{The rms units are arbitrary per data set.}
The solid horizontal lines show thresholds for discarding the worst
images (25\% for data~A, 33\% for data~B). 
The dashed lines show the mean rms for the remaining images. 
{\em Other panels:\/} corresponding image center cutouts
at the best seeing, seeing near the mean-above-threshold value, and
seeing near the threshold value, sampled at the times marked by three
squares in the graphs \reva{that are close for sampling the same
solar scene.} 
The three triangles in the upper graph mark the samples in the upper
row of Fig.~\ref{fig:overview}.
The three triangles in the lower graph mark the samples of the \PubB\
PHE, contrail fibril, and return aftermath shown in
Figs.~\ref{fig:overview} and \ref{fig:onions-samples}.
The two pairs of circles at the right mark 4 min delay pairs at good and
poor seeing used in Fig.~\ref{fig:onions-samples}.
The image greyscales along rows share the clipped byte scaling of the
sharpest images at the left, including saturation of the pore in the lower
row to boost granular contrast.
}
\end{figure*}

\section{Results}    \label{sec:results}
Our results consist of figures that we first describe one by one while
deferring their overall interpretation to Sect.~\ref{sec:interpretation}.

\paragraph{Figure~\ref{fig:overview}.}
We start with full-field samplings of data~A and B to show that both
data sets targeted patches of somewhat active network in quiet-Sun
areas.
\reva{Both were near disk center.}
The upper half of the data~A field was mostly network free and
resembles the quiet-Sun area in the upper left \reva{part} of the
field studied by
\citetads{2007ApJ...660L.169R}. 

The sample timings in Fig.~\ref{fig:overview} are specific
selections.
For data~A the first panel is at the moment with best seeing
(Fig.~\ref{fig:seeing}), with its central part enlarged in the second
panel of Fig.~\ref{fig:seeing} to show its quality.
\reva{The second and third are for reasonably good seeing moments
4 and 6~min later, respectively.}
For data~B the first panel shows the \PubB\ PHE, the second panel the
subsequent contrail fiber \reva{4~min later}, and the third panel the \revapar 
return aftermath \reva{6~min later}. 

The three lower panels represent a concise summary of \PubB. 
\reva{The visibility of} the striking PHE at left at the AIA \reva{304,
171, 193, and 94\,\AA\ wavelengths implies full} hydrogen ionization
\reva{because the corresponding formation temperatures are well above
the 20\,000~K value at which hydrogen ionizes fully in coronal
equilibrium (at lower temperature for higher density, see
\linkpubbpage{7}{Fig.~6} of \PubB)}.
Minutes later, the \Halpha\ core showed a fat dark fibril along its
track (second panel). 
Just below it, an adjacent one appeared as well, made by a preceding
shorter PHE called ``contrail B'' in \PubB.
Both retracted in their return aftermath with considerable \Halpha\
core redshift (third panel). 

Our quest here is to determine whether the less spectacular but more common
PHEs that are visible as RBEs produce smaller \Halpha\ core fibrils in a similar
manner.
Both data samples in Fig.~\ref{fig:overview} show RBE candidates in
the first column. 
Many seem to have corresponding darkenings in the 6 min delay third
column,
\reva{seen best with the blinkers in the online material.}

\paragraph{Figure~\ref{fig:seeing}} \hspace{-1.5ex} 
\revc{shows} the seeing during the two observation periods.
As usual, it varied considerably. 
We must account for these variations because we applied statistical
measures to the whole sequences in the form of temporal averaging and
scatter correlation, whereas the detection of slender RBEs and
aftermath features is very sensitive to image sharpness.

Figure~\ref{fig:seeing} shows our choice to discard the worst
quarter of all images for Data A in whole-sequence analysis and the
worst third for Data B, which had slightly lower quality, as evident in
the image samples.
The rightmost images show that even these no-pass thresholds still
correspond to image quality that may be deemed good at most other
solar telescopes.

\begin{figure*}
  \centering
  \includegraphics[width=\textwidth]{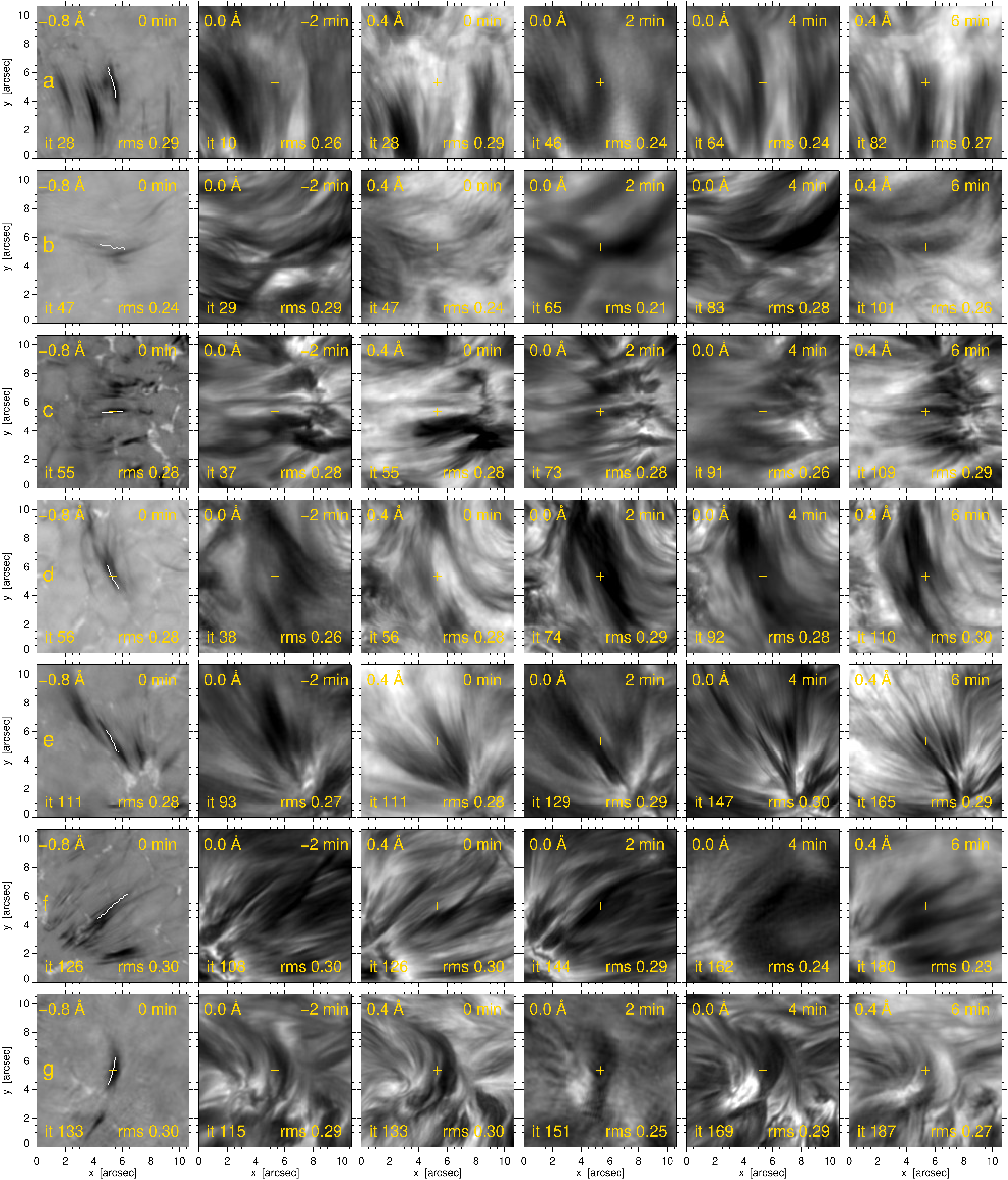}
  \caption[]{\label{fig:selecttracks} 
  Example sequences of RBEs and their aftermaths from data~A in the
  form of \Halpha\ image cutouts. 
  The images can be zoomed in per pdf viewer \reva{and the online movie versions can be inspected and enlarged.} 
  The $\Delta \lambda$ and $\Delta t$ values are specified at the top
  of each panel, the time steps (it) and seeing quality (rms) are at the
  bottom.
  {\em Rows\/}: seven cases $a-g$ in time order.
  {\em First column\/:} RBE skeleton taken from Pub\,A (white)
  overlaid on a cutout of the \Halpha\ blue-wing image at
  \reva{$\Delta \lambda \tis -0.8$~\AA\ } from mean line center. 
  The cutout is centered on the skeleton center marked with a plus.
  {\em Other columns\/}: \Halpha\ cutouts at $\Delta \lambda \tis 0.0$
  or $+0.4$~\AA\ from mean line center at time delays
  $\Delta t \tis -2, 0, 2, 4, 6$~min with respect to the the RBE
  measurement. 
  The byte scaling is defined per row and per wavelength by the image
  with the best rms value.
  \reva{The} \revapar online material \reva{includes three-panel
  movies per case} showing all pertinent time steps at the three
  sample wavelengths. \revb{Direct links per case:
  \online{figure03_case_a.mp4}{$a$},
  \online{figure03_case_b.mp4}{$b$},
  \online{figure03_case_c.mp4}{$c$},
  \online{figure03_case_d.mp4}{$d$},
  \online{figure03_case_e.mp4}{$e$},
  \online{figure03_case_f.mp4}{$f$},
  \online{figure03_case_g.mp4}{$g$}.
}}
\end{figure*}

\paragraph{Figure~\ref{fig:selecttracks}} \hspace{-1.5ex}
\revc{shows} examples of \PubA\ RBEs, denoted $a-g$ in the
seven rows. 
\revapar These seven cases were selected from the 608 RBE skeletons
detected in \PubA\ per visual SHOWEX inspection favoring good seeing
and RBE isolation for less confusion. 
Many others show yet more confused and confusing scenes, but the
cleaner examples in Fig.~\ref{fig:selecttracks} are already a good
demonstration of the utter richness, dynamic character, and small
scales that are typical for \Halpha\ fibrils around network. 
The online movie versions show such cutouts for each case at the three
selected wavelengths while covering all time steps during a
$[-2,+6]$~min range around the \PubA\ skeleton determination.

\begin{figure*}
  \centering
  \includegraphics[width=\textwidth]{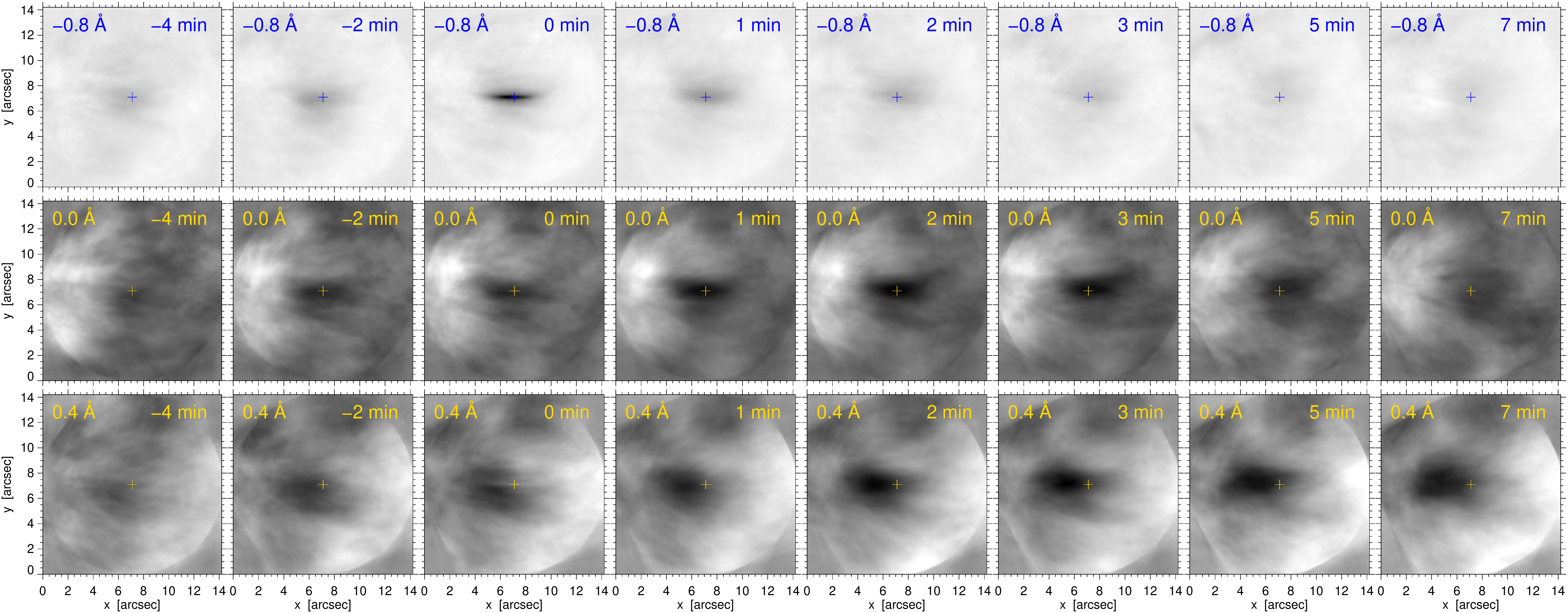}
  \caption[]{\label{fig:avertracks} 
  Stacked \Halpha\ image cutouts for data~A. 
  Before addition, all seeing-passed image cutouts are centered on the
  center of the RBE skeleton (plus) and rotated to point the
  skeleton to the right, away from the network to the left.
  {\em Rows\/}: averaged \Halpha\ samplings at
  $\Delta \lambda = -0.8, 0.0$ and 0.4\,\AA\ from mean line center.
  {\em Columns\/}: averaged samplings at different time delays with
  respect to each RBE track, from 4~min earlier sampling at the left to
  7~min later sampling at the right. 
  The $\Delta \lambda$ and $\Delta t$ values are specified at the top
  of each panel.
  The greyscales are shared along rows, defined by bytescaling the
  $\Delta t \tis 0, 2, 3$\,min panels, respectively.
  Blank corners result from cutout rotations.
  }
\end{figure*}

Each skeleton is shown by a white line at the centers of the blue-wing
image cutouts in the first column. \revapar
The skeleton midpoint was used as center for the cutout per case and
is marked with a plus, as a spatial reference 
also in the other panels per row and \reva{in} the other movie frames.

The blue-wing samplings are closer to line center than the
$\Delta \lambda \tis -1.3$~\AA\ used for the skeleton detections in
\PubA\ in order to show the RBEs fatter and longer around these.
When the same RBE was detected as a \PubA\ skeleton in multiple time
steps, we selected the RBE with the best seeing at its detection time
and the shown delay times.

The second column shows the line-center scene two minutes earlier.
Case $a$ shows a dark bunch of fibrils at the left that were likely
created just before and vanished with time (other panels).
At poor seeing (fourth panel) it seemed a thick homogeneous fibril, but
at the best seeing (third panel) it appeared finely striated. 
Case $e$ has a similar bunch at the skeleton position and timing.

The third column is simultaneous with the RBE in the first column and
also samples the \Halpha\ core, but not at line center but at
$\Delta \lambda \tis +0.4$~\AA\ to emphasize redward line shifts. 
It also displays the blueshifts that are part of the RBE phenomenon in
the form of cospatial brightenings.
These show up as thin bright stripes, thinner than the blue-wing RBEs
(the reason is discussed in Sect.~\ref{sec:discussion}).
They duplicate the skeletons in the first column in their location
(through the crosses) and orientation, but they extend \reva{farther}.
Such thin bright RBE stripes are seen in all seven panels of the third
column, but they are near the SST resolution limit and only visible at
the best seeing so that they are not a robust alternative to blue-wing
RBE detection. 
\reva{In these} best-seeing samples they suit better \reva{than the
skeletons} for comparison with the delay images to the right. 
These are at $\Delta t \tis 2, 4, 6$~min, the first two for line
center to check for contrail fibrils, the last one at
$\Delta \lambda \tis +0.4$~\AA\ to check for return aftermaths with
core redshifts as in \linkpubbpage{5}{Fig.~4} of \PubB.

We discuss the seven cases one by one.
\revb{In case~$a$ (\online{figure03_case_a.mp4}{movie}),} the 0 min panel
(third in the top row) has no dark fibril at its center, but two
bunches to the sides that may be related to previous RBE activity, as
suggested by other RBE locations and directions in the first panel. 
The three delay panels show a dark fibril, likely a striated bunch, but
the 2 min and 4 min seeing was less good. 
It has shortened in the 6 min panel, suggesting contraction, as in the
\PubB\ return aftermath.

In case~$b$ (\online{figure03_case_b.mp4}{movie}), the thin bright
stripe in the third panel suggests that the skeleton at the left is part
of the long curved thin gray feature.  
This obtains a dark fibril around it at 4 min delay, with a weaker and
shorter remnant at 6 min delay.

In case~$c$ (\online{figure03_case_c.mp4}{movie}), the thin bright
stripe in the third panel is part of a larger blueshift feature
pointing left that was also present two minutes before. 
It neighbors a dark fibril bunch below it that is already gone in the
2 min panel, possibly replaced by new (RBE?) blueshift.
The best indication of a fibril-after-RBE is in the 6 min red-core
panel, retracted to the right of the plus.

In case~$d$ (\online{figure03_case_e.mp4}{movie}), the thin bright
stripe in the third panel curves to the bottom right, followed by a
wider fibril bunch and replicated as a thin dark stripe in the
rightmost panel. 
The other nearby features illustrate how easily these scenes become
confused. 

Case~$e$ (\online{figure03_case_e.mp4}{movie}) already shows a fibril
around the plus in the second panel, which is mostly gone in the next
panel, then a dark bunch at 2 min that retracts at 4 min, and a long,
thin aftermath feature in the last panel.
There is suggestive correspondence between dark features in the 4 min
panel and the RBE panel at left. 

Case~$f$ (\online{figure03_case_f.mp4}{movie}) is again one in which
it seems best to compare the thin bright curved stripe in the third
panel with the fibrilar darkening in the later panels. 
The same holds for case~$g$ (\online{figure03_case_g.mp4}{movie}), in
which the curved bright stripe seems to become replicated as a dark
strand in the 4 min panel; it retracts in the last panel.

None of these cases is as obvious and straightforward as the PHE,
contrail fibril, and return aftermath of \PubB. 
However, they all suggest roughly similar behavior in their
development. When the dark features around the pluses in the first
and the other columns are compared, they seem to match better at 4 min delay than
between the simultaneous first and third columns. 
Yet better matches occur between the thin, bright RBE stripes in the
third column and slender dark features in the last column.
Lack of one-to-one dark-dark similarity between RBEs and line core
fibrils was already remarked in \PubA\ (caption to its
\linkpubapage{5}{Fig.~3}).
The examples in Fig.~\ref{fig:selecttracks} suggest that better
similarity is reached at multiple-minute delay in core imaging, but
with much competition from neighboring similar features in other
development stages. 

This is yet more evident in the online movie versions of these seven
cutout areas.
They vividly illustrate that at high angular and temporal resolution
the \Halpha\ chromosphere around network consists of a bewildering
multitude of fast-changing small features for which direct
cause-and-effect identification is much impeded by similar competitors
and also by intervals of worse seeing.

\begin{figure*}
  \includegraphics[width=\textwidth]{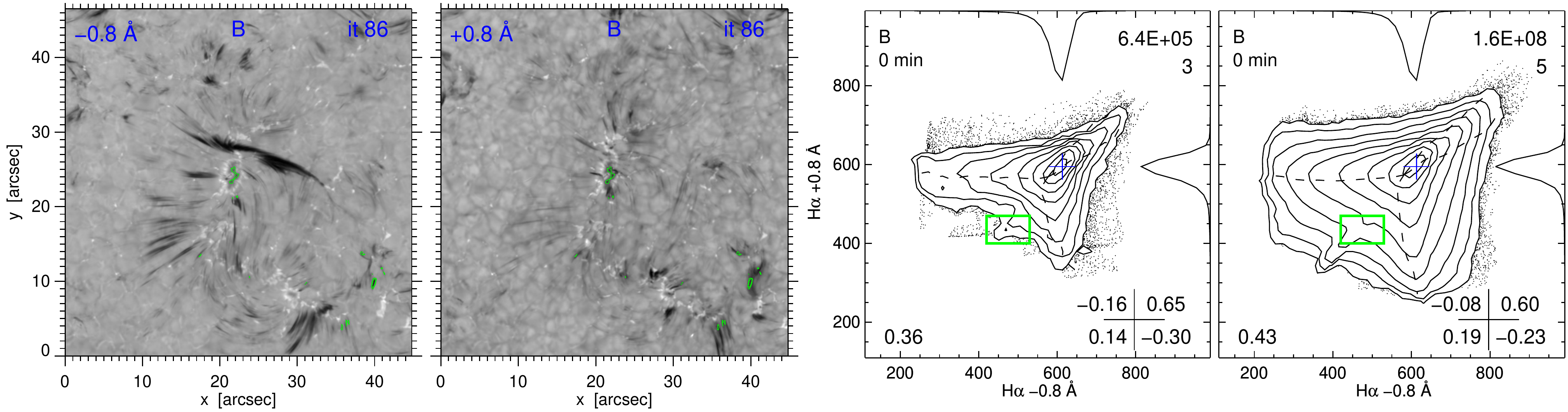}
  \caption[]{\label{fig:strousdemo} \reva{ Strous-format scatter
  analysis for data~B.
  {\em Images\/}: Simultaneous blue- and red-wing images at
  $\Delta \lambda \tis \pm0.8$~\AA\ at the best-seeing moment (highest
  square in the lower graph of Fig.~\ref{fig:seeing}). 
  {\em First diagram\/}: Scatter contours for this image pair. 
  The format is detailed in Sect.~\ref{sec:observations}.
  The axis quantities are intensities in SST data units.
  The green box selects a specific low-density contour feature with
  the corresponding pixels outlined by green contours in the images.
  {\em Second diagram\/}: Scatter contours for all such 252 image
  pairs in the data~B sequence passing the seeing threshold, with the
  same axes and the green box repeated as location reference.
  \revb{Online image blinker: \online{figure05_blink12.pdf}{1-2}}.
  }}
\end{figure*}

\paragraph{Figure~\ref{fig:avertracks}.}
In view of the small-scale confusion evident in our browsing with
SHOWEX and illustrated in Fig.~\ref{fig:selecttracks} and its movie
versions, we now turn to statistical full-duration analyses of the data
\revapar sequences, excluding their worst seeing samples \revapar
(25\% and 33\%, respectively).  
Figure~\ref{fig:avertracks} is made from the data~A \Halpha\ sequence
and the \PubA\ skeleton list by cutting out subimages centered on the
midpoint of each skeleton and rotating each subimage so that all
skeletons point to the right, emanating from network to the left.
All subimages passing the seeing threshold of Fig.~\ref{fig:seeing}
are then summed to obtain averages. 
The number of summed pairs varies between 200 and 300 depending on the
seeing sampling and on delay values exceeding the observing duration.

As expected, the top row shows RBEs averaging to a dark slender
horizontal feature at 0 min delay.
It is also weakly visible in the adjacent panels.
This presence might be attributed to RBE longevity, but the RBE
lifetime histograms in \linkpubapage{10}{Fig.~13} of \PubA,
\linkadspage{2012ApJ...752..108S}{6}{Fig.~4} of
\citetads{2012ApJ...752..108S}, 
and especially in \linkadspage{2013ApJ...764..164S}{5}{Fig.~4} of
\citetads{2013ApJ...764..164S} 
show that RBEs typically last only half a minute, rarely above a
minute.
Our SHOWEX inspections and Fig.~\ref{fig:delays} below suggest that this weak
presence is instead due to RBE recurrence, in agreement with
\citetads{2013ApJ...764..164S}, 
who found that on average, RBEs occur every 84\,s at locations around
network where at least one appeared during their observation.
Such fast RBE repetition contributes confusion in
Fig.~\ref{fig:selecttracks} and apparent longevity here.

The second row shows an elongated RBE signature in the 0~min panel
because RBEs tend to have core darkening (first column of
Fig.~\ref{fig:profiles}).
The $-2$~min panel shows a similar but vaguer feature, similar to the
top row, which we also attribute to frequent RBE recurrence.
Subsequently, darker elongated correspondence clouds with the same
horizontal orientation develop during $1-3$~min delay and then become
more diffuse.
These imply that statistically, there is significant probability for
fibril-shaped RBE-aligned core darkening after RBEs.
If this were not the case, these averaged image cutouts would instead
be smoothly gray or show vague blobs, as in the -4 and +7~min panels.

In the bottom row the thin, bright RBE stripes in the red-core images
in the third column of Fig.~\ref{fig:selecttracks} add up to produce a
bright thin horizontal feature in the third panel, longer and thinner
than the RBE feature in the third panel of the top row, but
representing the same phenomenon.
In the later samplings the elongated black blob darkens as in the
second row.
It extends left of the plus due to fibril retraction back to the
network root in return aftermaths. 

There are vaguer and more roundish blobs in the first and last columns
of all three rows, with the one in the last panel still quite dark and
similar ones in the $-2$ and 0~min panels of the bottom row spreading
around the thin bright RBE stripe in the 0~min panel.
We attribute these to close-lying companion RBEs at different times.
Our SHOWEX inspections, the examples in Fig.~\ref{fig:selecttracks}, and the
studies of Sekse \etal\
(\citeyearads{2012ApJ...752..108S}, 
\citeyearads{2013ApJ...764..164S}, 
\citeyearads{2013ApJ...769...44S}) 
show that RBEs are often launched close together with similar
orientation, for example the pair in
\linkadspage{2013ApJ...764..164S}{10}{Fig.~10} of
\citetads{2013ApJ...764..164S}. 
Frequent adjacency adds to frequent recurrence as darkener in these
image summations.

\reva{
\paragraph{Figure~\ref{fig:strousdemo}.}
We now turn to Strous-format scatter correlations. 
This figure introduces the format with a relatively straightforward
comparison from data~B in which well-known agents define the
solar scenes.
The images sample the outer \Halpha\ wings at equal wavelength
separation from line center ($\Delta \lambda \tis\, \pm\,0.8$~\AA),
simultaneously at the best-seeing moment.
Their blinker in the online material helps to appreciate their
correspondences and differences.

The first diagram shows the corresponding Strous-format scatter
correlation.
The second diagram results from treating all instantaneous image pairs
in the data~B sequence above the seeing-quality threshold of
Fig.~\ref{fig:seeing} likewise.
The increased statistics then give smoother contours that sample
steeper slopes and spread farther from the summit.

The main agents sampled in the outer \Halpha\ wings are the
granulation in the deep photosphere, which covers most of these fields
and dominates the summit contours, magnetic bright points, which result
in the narrow bright-bright mountain ridge pointing to the upper
right, and RBEs in the blue wing and RREs in the red wing, which cause the
wider ridges from the summit leftward and downward. 
Lesser agents such as the $p$-mode Dopplershift pattern cause further
blurring.
There is no signature of\ reversed granulation, gravity waves,
internetwork shocks, magnetic canopies or other upper-photosphere or
chromospheric structures because \Halpha\ has an extinction gap
between the deep photosphere and the chromosphere
(\citeads{1972SoPh...22..344S}; 
\citeads{2012A&A...540A..86R}) 
and senses the latter only closer to line center in other fibrils than
RBEs and RREs (including the long ones studied here).

The appearance of the granulation in the two images differs through the
intensity-Dopplershift correlation (bright granules rise,
dark lanes subside), which doubly shifts each wing accordingly
through the fixed passband. 
This asymmetry diminishes granular contrast in the blue wing, enhances
it in the red wing, and defines the shape of the mountain summit with
near-parallel moment curves.
It also enhances bright-point contrast in the blue wing
(\citeads{2006A&A...449.1209L}); 
additional blue-wing bright-point enhancements result from commonly
occurring downdrafts in magnetic concentrations
(\citeads{2010ApJ...709.1362L}). 
They tilt the bright-point scatter ridge clockwise from the diagonal
direction.  
It widens slightly in the full-sequence diagram, with a lower
upper-right quadrant Pearson coefficient, because all additional images are
less sharp.

The \PubB\ PHE in the first image contributes heavily to the first
diagram with very dark blue-wing pixels, but no signature in the red wing.
However, its contribution to the second diagram is negligible because
it lasted only briefly; when its time segment is removed from the
summation, the contours and coefficients remain virtually identical.
In this diagram the leftward ridge is wider than the downward ridge
because there are more RBEs than RREs
(\citeads{2013ApJ...769...44S}). 
RBEs also reach darker extremes.
These features are mutually exclusive because the required core
Dopplershift is blueward for RBEs, redward for RREs, and makes these
chromospheric features transparent in the other wing so that it
samples the granulation far underneath.
The darkest pixels in one sample the distribution for the other
without preference; the moment curves return to the
distribution-maximum locations.

The green box in the first diagram emulates SHOWEX identification of
which pixels cause what scatter-diagram feature. 
The corresponding green-contoured pixel pairs in the images show that
most of the unusual-looking extension of the outer contour came from
the single pore near image center. 
It covered only a promille of the field so that it shows up only
beyond the sixth contour from the summit ($3^{-6} \tis\ 0.0013$).
Through its longevity during the sequence, it gained a much clearer
signature in the time-summed diagram at the right: a conspicuous narrow
high-correspondence ridge with 45\degree\ response equality.
Masking the pore produced the same diagram without this ridge.
Its presence would remain the same if the pore had moved around or had
appeared at a random location per image.

This example of Strous-format scatter diagram production demonstrates
that the full-sequence variant is useful to show and disentangle
statistical properties for multiple agents that occur briefly but
frequently (such as granulation, RBEs, and RREs) or more persistently (such as
magnetic bright points and the pore), even at small or minute filling
factor (such as the pore).

In this multi-agent comparison the linear overall Pearson correlation
coefficient is a useless estimator. 
The quadrant coefficients do better but remain too biased toward the
summit.  
This is evident for the lower left one, which has relatively high
values from the initial correspondence near the summit where the
moment curves cross before becoming nearly no-correlation Cartesian.  
Masking the pore reduces the full-sequence lower left value only from
0.19 to 0.18 and does not affect the moment curves.
The other two quadrant coefficients are also set by the tilt of the
summit contours.
Inspection of the shape of the low-density outer contours is much
better for recognizing correlation signatures of different
nondominating but persistent or recurring agents, as demonstrated by
the pore.

Finally, this figure suggests that RBEs and RREs might be used in tandem to increase
spicule II statistics, for example by selecting the darkest pixels in
either wing per time step. 
However, in many figures we compare RBE presence to later core
redshift in searching aftermaths with return flows.
Because RREs also have core redshifts, ambiguities would then arise
between them and their aftermaths. 
We therefore restrict most figures to RBEs only.

\begin{figure}
  \includegraphics[width=88mm]{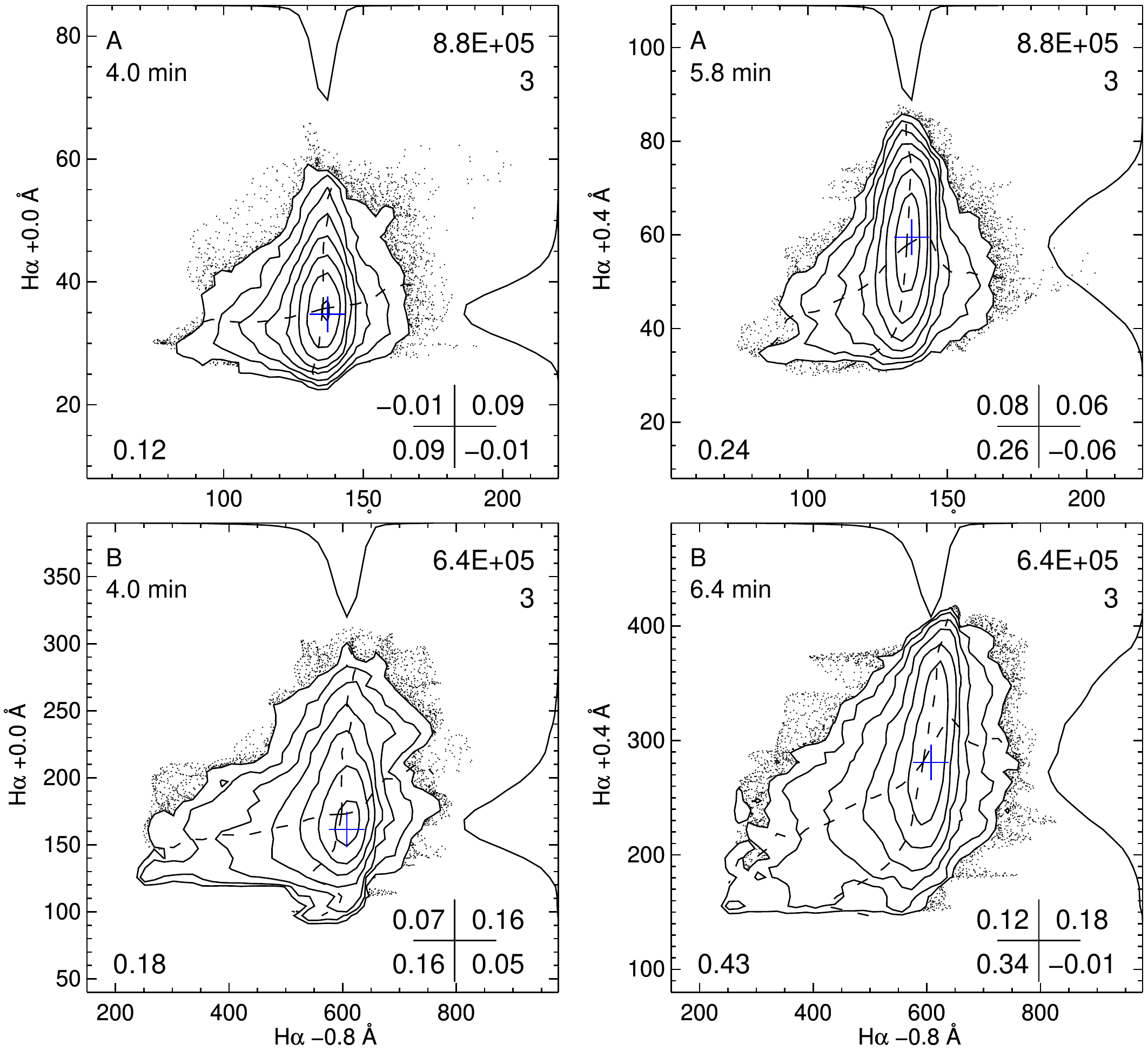}
  \caption[]{\label{fig:onions-images} \reva{
  \revapar Scatter diagrams for the data~A ({\em upper row\/}) and
  data~B ({\em lower row\/}) image pairs in Fig.~\ref{fig:overview}.
  {\em Left column\/}: intensity at the line center of \Halpha\ at
  4 min time lag against intensity in the blue wing at
  $\Delta \lambda \tis -0.8$~\AA\ 4~min.
  {\em Right column\/}: intensity at $\Delta \lambda \tis +0.4$~\AA\
  at about 6 min lag against intensity in the blue wing.
  }}
\end{figure}

\paragraph{Figure~\ref{fig:onions-images}} \hspace{-1.5ex}
\revc{initiates} \revc{our} time-delay scatter analysis by employing it to
the six images in Fig.~\ref{fig:overview}.  
We there remarked that their inspection suggests correspondence
between RBE presence in the first column and later red-core darkening
in the third column.  
Here this apparent correspondence manifests itself as extensions
(``spurs'') of the contour mountain left and down from the summit. 

The first contours from the summit suggest no correspondence except
for the slight tilts in the last diagram.
Figure~\ref{fig:overview} shows that this lack of correspondence is
dominated by the very quiet internetwork sampled especially in the
upper part of data~A. 
It has no RBEs but only granulation in the \Halpha\ blue wing, and it
is poor in fibrils in the \Halpha\ core. 

The outer contours instead show left-down mountain spurs.
In data~B the Pub~B PHE, contrail, and aftermath contribute to these,
\revapar making them more pronounced with higher lower-left Pearson
quadrant coefficients.
The spurs are most pronounced in the red-core diagrams at the right and
show large dark-dark correspondence.
For example, in the final diagram, all RBE pixels darker than 400 data
units were also darker than average 6.4~min later in the red-core
image, with the darkest followed by darkest. 
However, there were also about as many darkest red-core pixels with
slight or no preceding wing darkening, and in the left-hand diagrams
there are yet darker line-center pixels without preceding wing
darkening.
Below we identify RREs and dynamic fibrils as contributors to these. 
} 

\begin{figure*}
  \includegraphics[width=\textwidth]{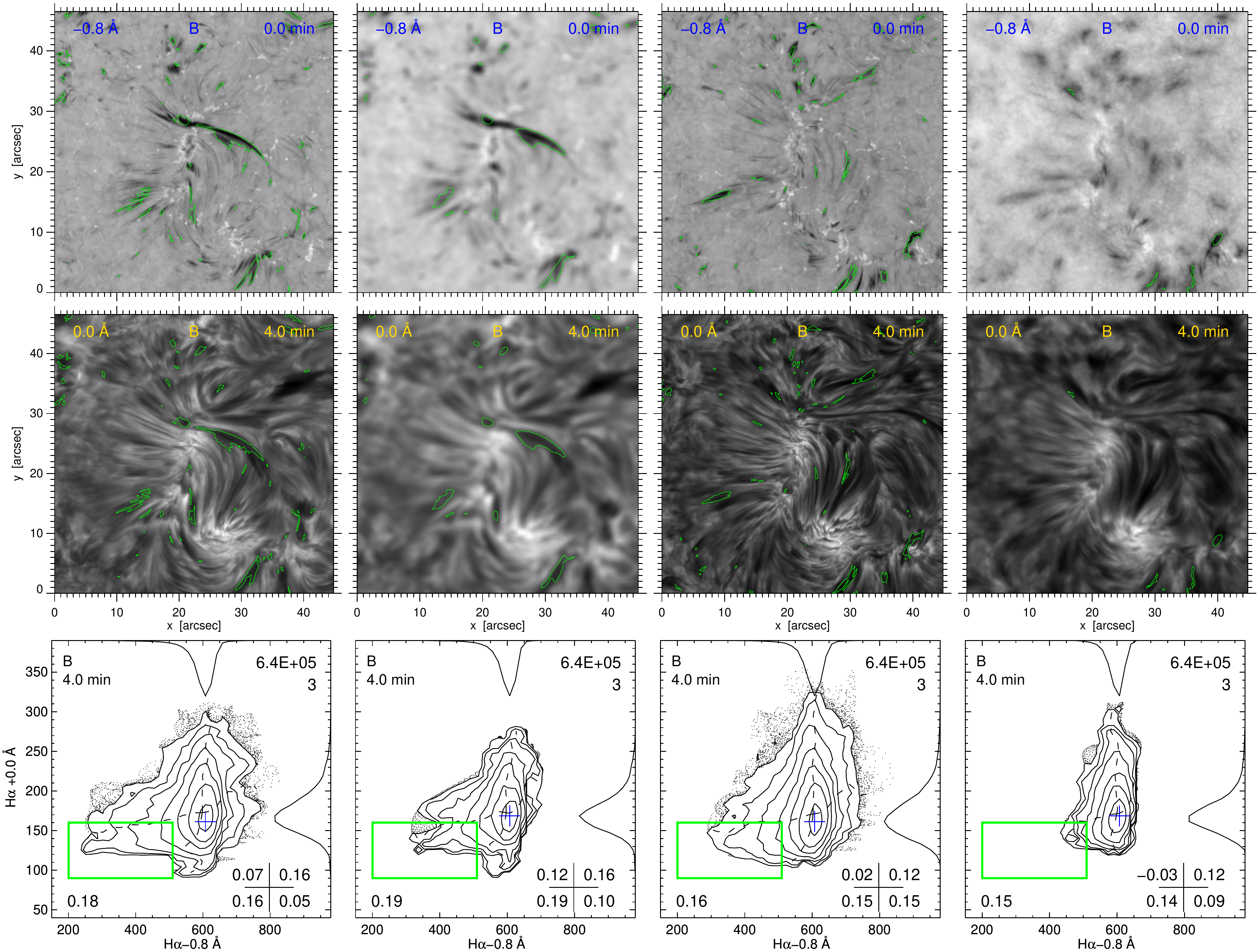}
  \caption[]{\label{fig:onions-samples} 
  Time-delay scatter analysis at good and poor resolution in data~B.
  {\em First row\/}: images in the blue wing of \Halpha\ at
  $\Delta \lambda \tis -0.8$~\AA. 
  \revapar
  {\em Second row\/}: images at \Halpha\ line-center taken 4~min
  later.
  {\em Third row\/}: \revapar scatter diagrams for each image pair
  \reva{above with} the same axes as in \reva{Figs.~\ref{fig:onions-images}
  and} \ref{fig:onions}. 
  \revapar The green box selects the lower left part with
  corresponding pixels outlined by green contours in the images
  above.  
  {\em First column\/}: \revapar image pair for the \PubB\ PHE and
  contrail, the same as in Fig.~\ref{fig:overview}.
  {\em Second column\/}: the same image pair smeared over 1~arcsec.
  {\em Third column\/}: \revapar image pair with good seeing.
  {\em Fourth column\/}: \revapar image pair with poor seeing.
  \reva{The latter pairs are} marked with circles \reva{to the right} in the
  lower graph of Fig.~\ref{fig:seeing}.  
  \revapar }
\end{figure*}

\paragraph{Figure~\ref{fig:onions-samples}} \hspace{-1.5ex}
\revc{shows} \reva{the effect of degraded resolution on scatter
diagrams.}
The images in the first column are the same as the first data~B images
in Fig.~\ref{fig:overview}, sampling the \PubB\ PHE and contrail
fibril, with the scatter diagram underneath repeating the lower left
diagram in Fig.~\ref{fig:onions-images} for reference.

The green box \revapar selects the spur part of the contour mountain
\reva{in the first diagram}, with the corresponding pixels outlined by
green contours in the images above. 
All belong either to the \PubB\ PHE or to RBEs pointing away from the
network. 
This part of the first scatter diagram therefore represents RBE
signatures including the \PubB\ PHE. 
\revapar

The second column treats the same images, but with 1~arcsec smearing.  
The contour spur shrinks considerably toward the mountain summit. 
The image contours show correspondingly that far fewer RBEs now
qualify as part of the scatter spur.

The third and fourth columns are similar demonstrations for the pair
of good-seeing images and the pair of poor-seeing images marked by
circles \reva{to} the right in the lower graph of
Fig.~\ref{fig:seeing}.
At each wavelength, the samplings were within 1.5~min 
of each other \reva{so that the scenes} differ primarily in
resolution. 
In the good-quality third-column panels the \reva{green box} \revapar
selection again selects RBEs (now without the extraordinary \PubB\
features), as shown by the contours in the images \reva{above,} which
\reva{mostly} display elongated shapes. 
In the \reva{final} poor-seeing pair the RBE \reva{selection} again
diminishes drastically \revapar and the scatter spur shrinks much
toward the contour mountain summit. 

\begin{figure*}
  \centering
  \includegraphics[width=\textwidth]{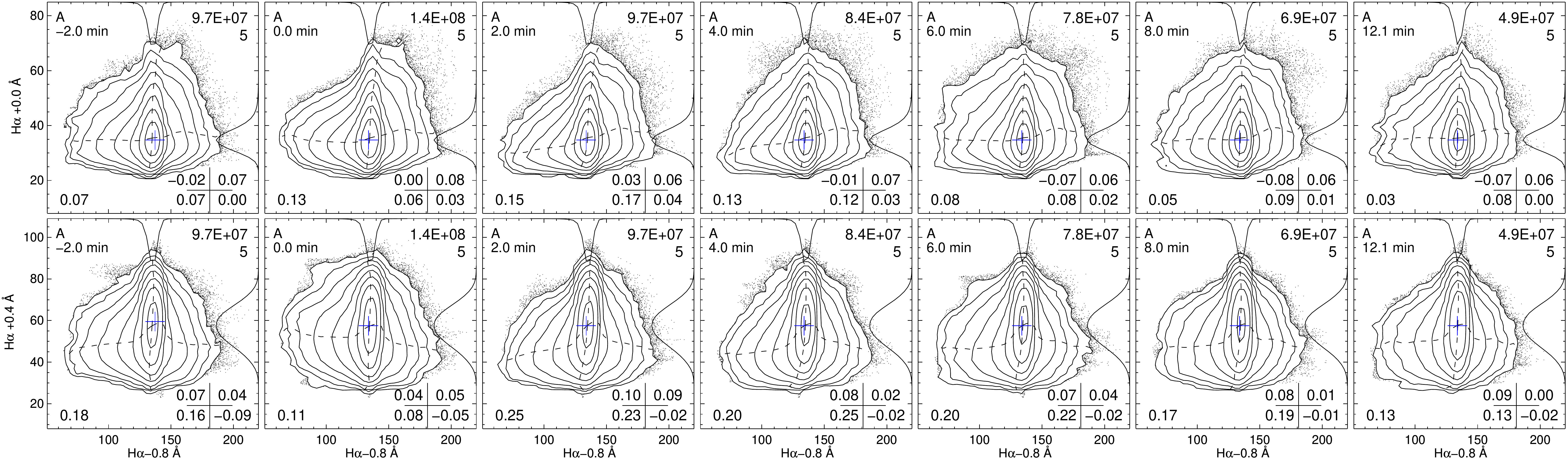}
  \includegraphics[width=\textwidth]{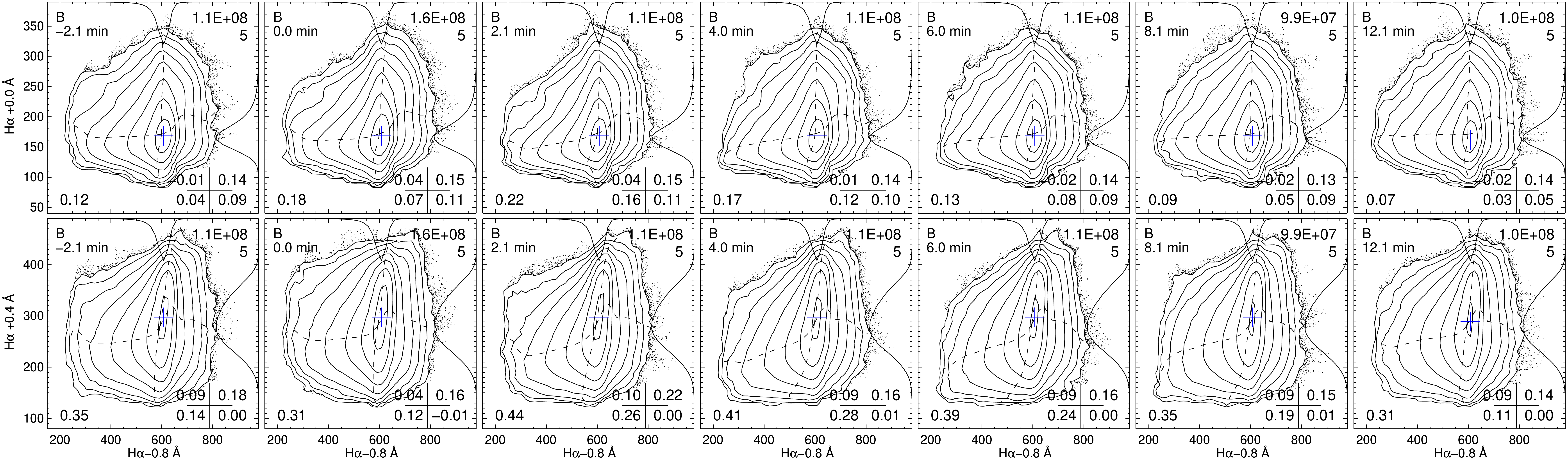}
  \caption[]{\label{fig:onions} 
  Full-sequence time-delay scatter \reva{analysis for data~A ({\em
  upper two rows\/}) and data~B ({\em lower two rows\/}).}
  \revapar In each pair the upper row shows the pixel-by-pixel
  correlation for \Halpha\ line-center intensity before or after the
  \Halpha\ blue wing intensity at $\Delta \lambda \tis -0.8$~\AA.
  The lower row \reva{per pair} shows the correlation of red-core
  intensity at $\Delta \lambda \tis +0.4$~\AA\ with blue-wing
  intensity.
  The sampling delays (positive for the core after the blue wing) are
  specified at the upper left in each panel.
  \revapar \reva{The worst-seeing images are discarded (25\% for
  data~A, 33\% for data~B).
  Axes are the same as Figs.~\ref{fig:onions-images} and
  \ref{fig:onions-samples}.}
  \revapar }
\end{figure*}

\paragraph{Figure~\ref{fig:onions}} \hspace{-1.5ex}
\revc{shows} scatter correlations combining image pairs from the
\reva{full} data sequences, excluding the worst-seeing samplings.
The many samples produce much better statistics than in
\reva{Fig.~\ref{fig:onions-images},} with \revapar \reva{well-defined}
contours reaching out \reva{farther} from the mountain summits,
\reva{as in Fig.~\ref{fig:strousdemo}}.

The first and third rows for Data A and B, respectively, chart
\Halpha\ line-center ($\Delta \lambda \tis 0.0$~\AA) intensity against
blue-wing intensity at $\Delta \lambda \tis -0.8$~\AA\ at different
time delays.
The second and fourth rows chart \Halpha\ red-core intensity at
$\Delta \lambda \tis +0.4$~\AA\ against blue-wing intensity.

The data~B $\Delta t \tis +4$~min diagram in the third row (central
panel) is the full-sequence version of the \reva{corresponding}
scatter diagrams in \reva{Figs.~\ref{fig:onions-images} and}
\ref{fig:onions-samples}. 
It shows the same mountain spur to the lower left, implying
\reva{significant} dark-dark \reva{association}. \revapar
The box selections in \revapar Fig.~\ref{fig:onions-samples}
suggest that most or all of these belong to RBE-like features.
The spur starts already at the mountain summit, 
\revapar implying that very many pixels have slightly dark -- slightly
dark \reva{association}.

The first two and last two panels of the third row do not show the
spur and show \reva{nearly} perpendicular moment curves, except for
the small bright-bright bulge in the horizontal one, which is
constantly present and \reva{probably} represents the network itself
(Fig.~\ref{fig:overview}).
The main change with delay duration in the third-row diagrams is the
downward tilt of the leftward mountain spur during delays
$\Delta t \tis 2-6$~min. 

The contour patterns and development in the first row for data~A are
similar to those for data~B in the third row.
The delayed downward tilt of the leftward spur corresponds to the
darkening blob in the central row of Fig.~\ref{fig:avertracks}.

\reva{The second row with sampling at $\Delta \lambda \tis 0.4$~\AA\
corresponds to the bottom row of Fig.~\ref{fig:avertracks}.
Compared to the top row, the onion-like contour patterns show an upward
shift of the mountain summits due to the very quiet areas in data~A,
as noted for Fig.~\ref{fig:onions-images}.}
The left- and downward spur is clearer in the $\Delta t \tis 2-4$~min
panels of the upper row, but in the second row, it persists to longer
delay, \revapar in agreement with the longer persistence of the
corresponding dark blob in Fig.~\ref{fig:avertracks}.

The fourth row shows the $\Delta \lambda \tis 0.4$~\AA\
delay samples of data~B, which are richer in blue-wing RBEs and core
fibrils (Fig.~\ref{fig:overview}). 
At 6 min delay, the lower left bulge even shows a promontory that
persists at 8 min delay. 
A bulge signature remains even at 12 min delay.

The two data~B rows start with a squarish contour rise at the upper left,
implying bright at $\Delta \lambda \tis +0.4$~\AA\ for dark RBE
locations.
Based on SHOWEX inspections, we attribute this to the thin bright
RBE stripes seen in the simultaneous data~A samplings in
Figs.~\ref{fig:selecttracks} and \ref{fig:avertracks}.

\begin{figure*}
  \centering
  \includegraphics[width=\textwidth]{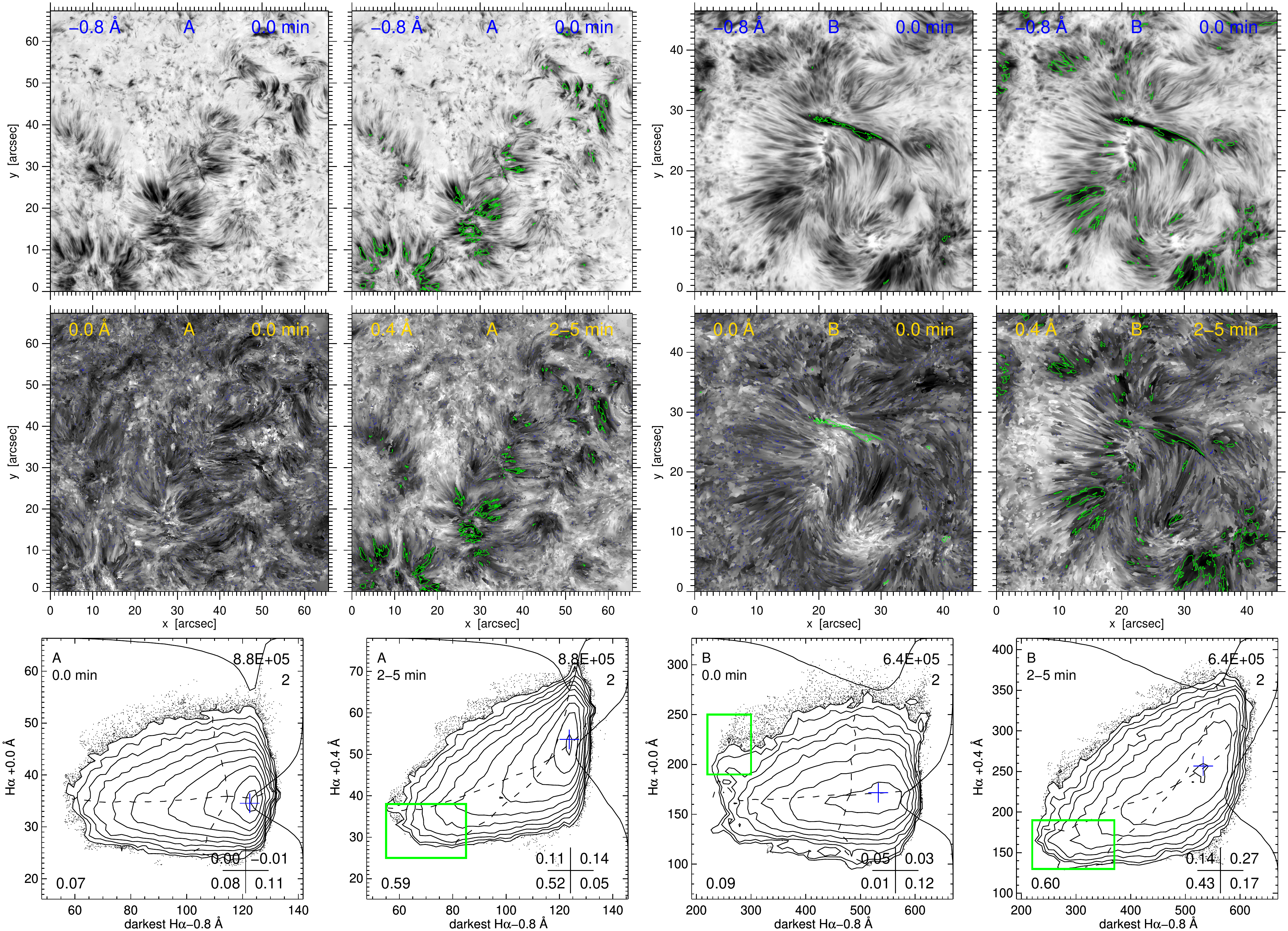}
  \caption[]{\label{fig:darkest} 
  Time-delay scatter analysis for the darkest RBE moments \reva{in
  data~A ({\em left two columns}) and data~B ({\em right two
  columns}).}
  {\em First row\/}: \revapar \Halpha\ images at
  $\Delta \lambda \tis -0.8$~\AA\ constructed by selecting  the darkest value for each
  pixel throughout the entire observation duration.
  The first pair of images \revapar is identical except for the green
  overlays; so is the second pair. \revapar {\em Second row\/}:
  simultaneous and delayed \Halpha\ line-core images. 
  {\em Third row\/}: scatter diagrams for each image pair above it. 
  {\em First and third columns\/}: data~A and B image pairs with the
  lower image made up of the same pixels at \Halpha\ line center at
  the same sampling times.
  {\em Second and fourth columns\/}: data~A and B image pairs with the
  lower image made up of the same pixels, but selecting the darkest
  value per pixel at $\Delta \lambda \tis +0.4$~\AA\ during the $2-5$~min
  delay range after the upper-image pixel sampling times.
  \revapar The green boxes select \revapar \reva{specific correlation
  features with} corresponding contours in the image pairs.
  \revb{Online image blinkers: 
  \online{figure09_blink15.pdf}{1-5}, 
  \online{figure09_blink16.pdf}{1-6}, 
  \online{figure09_blink37.pdf}{3-7},
  \online{figure09_blink38.pdf}{3-8}}.
}
\end{figure*}

\paragraph{Figure~\ref{fig:darkest}.}
Because Fig.~\ref{fig:onions} exhibits pixel-by-pixel correlations
without information on feature membership, in particular whether the
dark-dark \reva{associations} indeed represent fibrils, this figure
recovers these signatures by assembling image constructs.
The first two columns show data~A, the third and fourth show data~B. 
The upper row shows pairs of identical images in which each pixel is
the darkest value at $\Delta \lambda \tis -0.8$~\AA\ for the entire sequence duration.  
There is no need to discard poor seeing because its contrast blurring
already disqualifies these images.
The second row alternatively shows the intensity at line center at the
same time and the darkest value at $\Delta \lambda \tis +0.4$~\AA\
during the $2-5$~min delay after the above pixel sampling. 
Selecting the darkest during this delay range also favors samples with
better seeing.
The bottom row shows corresponding scatter diagrams; \reva{three have
green selection boxes of specific contour features}.

Random patterning might be expected for this darkest-pixel selection,
the more so for data~B with its three times longer duration and
increased activity, but the sharp fibrilar detail in both the first
(data~A) and third image (data~B) shows that these very
darkest pixels are also arranged in fibrilar alignments. 
This suggests that adjacent darkest pixels either sample the same RBE
or successive ones that follow similar tracks at different times. 
Below we show that both types contribute, and also that subsequent RBE
samples tend to be close in time (Fig.~\ref{fig:delays}).
The first image furnishes RBE mappings that extend well beyond the
single-snapshot skeletons shown in Fig.~\ref{fig:selecttracks}.
In the third image the \PubB\ PHE is prominently present over its full
area covered in time.
Thus, the blue-wing constructs represent renderings of the darkest
parts of RBEs during the observing periods.

The delayed red-core images in the second row show high correspondence
with the top row images, much more than the simultaneous line-center
images.  \reva{This can be seen in the online panel blinkers
and} is quantified by the correlation diagrams in the bottom row.
The second and fourth are similar to the corresponding rows in
Fig.~\ref{fig:onions}, but show tighter dark-dark \reva{association} in
the \reva{left-and-down pointing} mountain spurs, \reva{with much
higher lower-left quadrant Pearson coefficients}. 
In contrast, at simultaneous sampling (first and third diagrams), there
is no correlation between RBEs and dark line-center fibrils; the
difference between the scatter mountains per pair is
striking. \revapar

The green-box selection of the darkest-darkest pixel pairs in the
\reva{second and fourth diagrams} shows that \reva{the corresponding
features} are indeed fibrilar in nature and are preferentially located
at RBE feet (top images).  
\reva{The green box in the third diagram selects an upward contour
promontory in which darkest-wing pixels are associated with
brighter-than-average line-center pixels. 
These correspond to} brightening by core blueshift in the launch
\reva{phase} of the \PubB\ PHE. \revapar

\begin{figure}
  \centering
  \includegraphics[width=88mm]{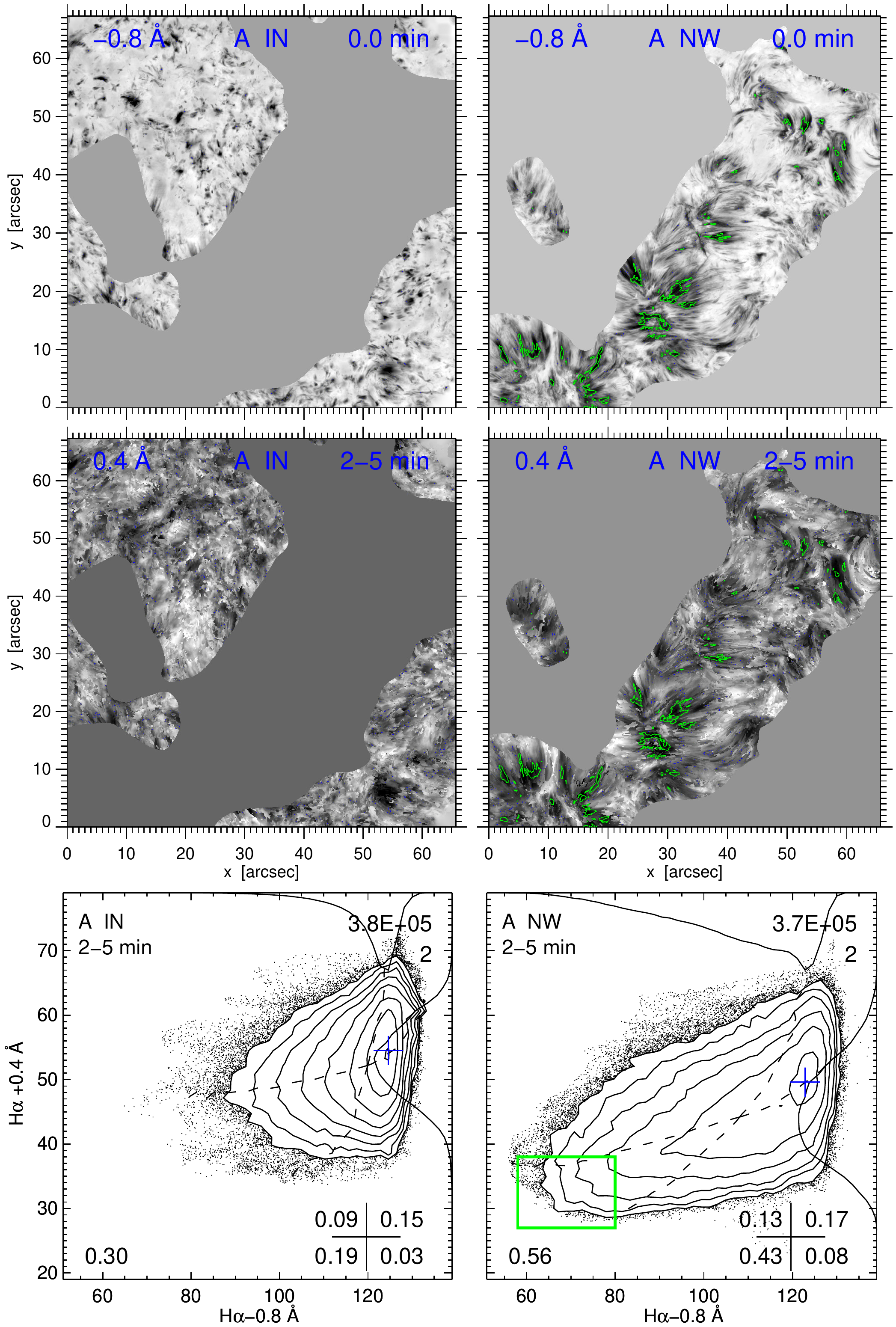}
  \caption[]{\label{fig:masked} 
  Time-delay scatter analysis for the darkest RBE moments in data~A
  \reva{only,} in the same way as in \reva{the first columns of}
  Fig.~\ref{fig:darkest}, but \reva{selecting inter}network \reva{({\em
  left\/}, label IN)} and network \reva{({\em right\/}, label NW)}. 
  \revapar \reva{{\em Upper images\/}:} darkest \reva{value per pixel}
  at $\Delta \lambda \tis -0.8$~\AA.
  \reva{{\em Lower images\/}:} darkest value per pixel at
  $\Delta \lambda \tis +0.4$~\AA\ during $2-5$~min delay after the
  upper-image pixel sampling. 
  \revapar \reva{{\em Scatter diagrams\/}: Corresponding}
  pixel-by-pixel correlations.
  \revapar The green box and contours \revapar select darkest-RBE and
  darkest delayed red-core pixels in the network areas.
  \revb{Online image blinkers: 
  \online{figure10_blink13.pdf}{1-3}, 
  \online{figure10_blink24.pdf}{2-4}}.
  }
\end{figure}

\paragraph{Figure~\ref{fig:masked}.}
The left-and-down mountain spurs in the time-delay scatter diagrams in
Figs.~\ref{fig:onions-samples}--\ref{fig:darkest} extend from close to
the summits of the contour mountains, suggesting that there are very many
pixels with slightly dark -- slightly dark \reva{association}. 
Figure~\ref{fig:masked} represents a test of their nature, in
particular whether they are contributed by quiet internetwork areas
without RBEs and fibrils.

The field of data~A is best suited to this test because it has wide
areas of very quiet fibril-free chromosphere toward its upper left
and lower right (Fig.~\ref{fig:overview}).
We constructed internetwork and network masks by setting brightness
thresholds on a heavily smeared (200~pixel boxcar) average over the
whole data~A sequence at $\Delta \lambda \tis +0.4$~\AA\ because the
third panel of Fig.~\ref{fig:overview} shows that this is a suitable
divider.

The scatter diagrams in the bottom row of Fig.~\ref{fig:masked} show
minor dark-dark \reva{association} for the internetwork area, whereas the
network diagram is closely similar to the second diagram in the bottom
row in Fig.~\ref{fig:darkest}, but with less upward spread above the
summit that was contributed by the internetwork, as seen in the first
diagram.
Hence, most pixels contributing slightly dark -- slightly dark
\reva{association} in the spur belong to the network areas, just as
the pixels with darkest-darkest \reva{association}. 
\reva{The latter are also darker in both diagnostics than any
internetwork pixel.}

\begin{figure}
  \centering
  \includegraphics[width=88mm]{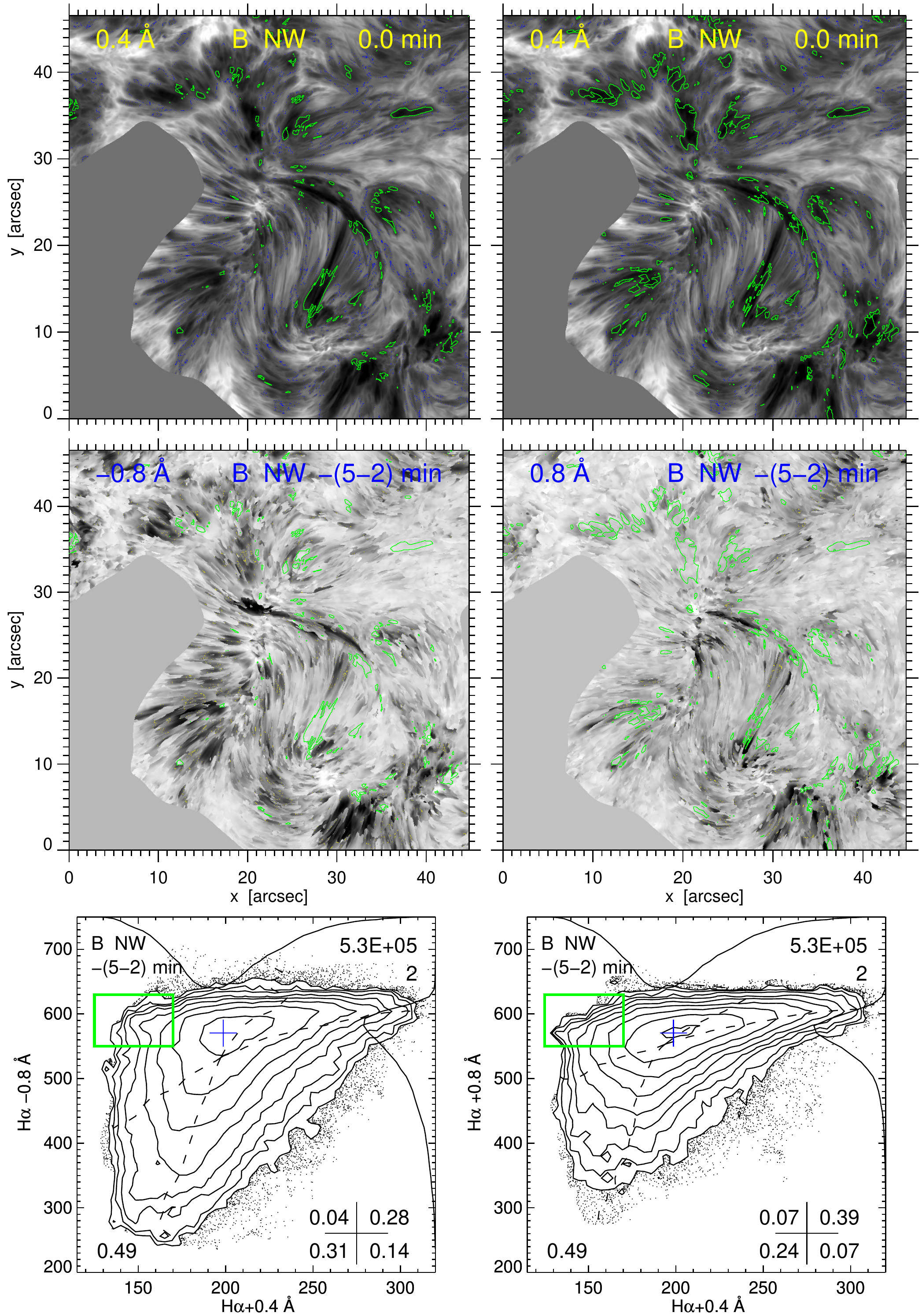}
  \caption[]{\label{fig:reversed} 
  Time-delay scatter analysis in reverse of Fig.~\ref{fig:masked} for
the  data~B \revapar network only.
  \reva{{\em Upper images\/}:} darkest-pixel core samplings at
  $\Delta \lambda \tis +0.4$~\AA. 
  \revapar Only the green contours differ.
  \reva{{\em Lower images\/}:} \reva{at the left we show} the darkest value per
  pixel at $\Delta \lambda \tis -0.8$~\AA\ \reva{to select preceding
  RBEs} during $5-2$~min before the upper-image pixel sampling times,
  \reva{and at the right} we show this at $\Delta \lambda \tis +0.8$~\AA\ \revapar to
  select \reva{preceding} RREs.
  \revapar \reva{{\em Scatter diagrams\/}: corresponding}
  pixel-by-pixel correlations.
  \revapar The green boxes \reva{and contours} select darkest red-core
  pixels without preceding blue-wing or red-wing darkening,
  respectively.
  \revb{Online image blinkers: 
  \online{figure11_blink13.pdf}{1-3} and 
  \online{figure11_blink24.pdf}{2-4} without contours, 
  \online{figure11_blink34.pdf}{3-4} with contours}. 
  }
\end{figure}

\paragraph{Figure~\ref{fig:reversed}.}
Figures~\ref{fig:darkest} and \ref{fig:masked} tested whether darkest
core pixels follow on darkest blue-wing pixels.
We now reverse this question into asking instead to what extent any
darkest core pixels relate to preceding RBEs. 
We also add red-wing RREs to this question. 

The identical images in the top row of Fig.~\ref{fig:reversed}
represent the darkest moment per pixel at
$\Delta \lambda \tis +0.4$~\AA\ during the whole sequence for network
areas only and for data~B because this has more network and showed
more RBEs and RREs than data~A. 
The network-only mask was constructed similarly to
Fig.~\ref{fig:masked}.

The first image of the second row displays the darkest blue-wing value
per pixel at $\Delta \lambda \tis -0.8$~\AA\ during the $5-2$~min time
range prior to the above pixel sampling. 
This image charts preceding RBEs.
The right-hand image similarly displays the darkest red-wing value at
$\Delta \lambda \tis +0.8$~\AA\ during $5-2$~min before to chart
preceding RREs.  
Both images display a multitude of fibrilar features, more and denser
in the left-hand image.
For each there are correspondences with red-core fibrils in the upper
image that are \reva{best seen in the column panel blinkers provided in the
online material}. 

The scatter diagrams in the bottom row are now triangular, implying
that the darkest red-core pixels have about as much probability to
follow on dark RBE pixels at the left or RRE pixels at the right as to follow
on non-dark pixels at these outer-wing wavelengths. 
Some of the latter no-correlation pixels are selected with green boxes
and marked by green contours in the images. 

\begin{figure*}
  \centering
  \includegraphics[width=\textwidth]{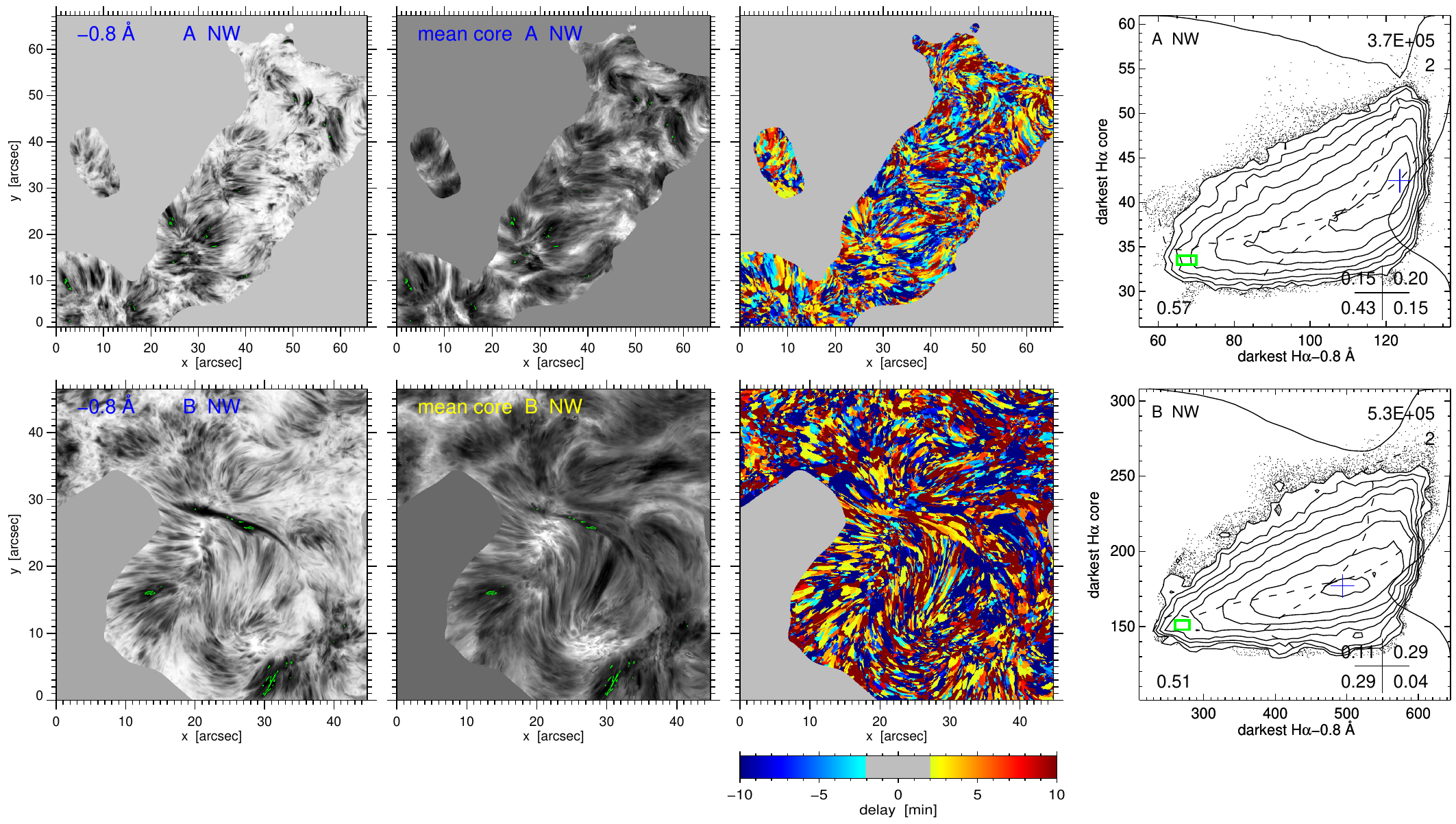}
  \caption[]{\label{fig:darkdark} 
  Sequence-wide dark-dark comparisons for \reva{data~A ({\em upper
  row\/}) and data~B ({\em lower row\/})} network only.
  \revapar {\em First column\/}: darkest value per network pixel
  during the whole sequence at $\Delta \lambda \tis -0.8$~\AA.
  The lower image is the same as the third image in
  Fig.~\ref{fig:darkest}, except for the mask and contours.
  {\em Second column\/}: darkest value per network pixel during the
  whole sequence of core intensities averaged over
  $\Delta \lambda \tis [-0.4,+0.4]$~\AA\ \reva{beyond} [-2,+2]~min
  \reva{delay} range around the blue-wing sampling per pixel. 
  {\em Third column\/}: color-coded time delay between darkest wing
  \reva{occurrence} and darkest core \reva{occurrence} per pixel,
  yellow to red for \reva{increasing positive delay (core after wing)
  beyond 2~min}, and light \revapar to dark blue for increasing
  \reva{negative delay (core before wing) beyond 2~min}.
  {\em Fourth column\/}: scatter \revapar diagrams between the first
  and second images per row.   
  The small green boxes and \reva{corresponding} contours select
  darkest-darkest pairs of which the \Halpha\ profiles are shown in
  Fig.~\ref{fig:profiles}. 
  \revb{Online image blinkers: 
  \online{figure12_blink12.pdf}{1-2}, 
  \online{figure12_blink13.pdf}{1-3}, 
  \online{figure12_blink23.pdf}{2-3},
  \online{figure12_blink45.pdf}{4-5}.
  \online{figure12_blink46.pdf}{4-6},
  \online{figure12_blink56.pdf}{5-6}}.
  }
\end{figure*}

\begin{figure}
  \centering
  \includegraphics[width=88mm]{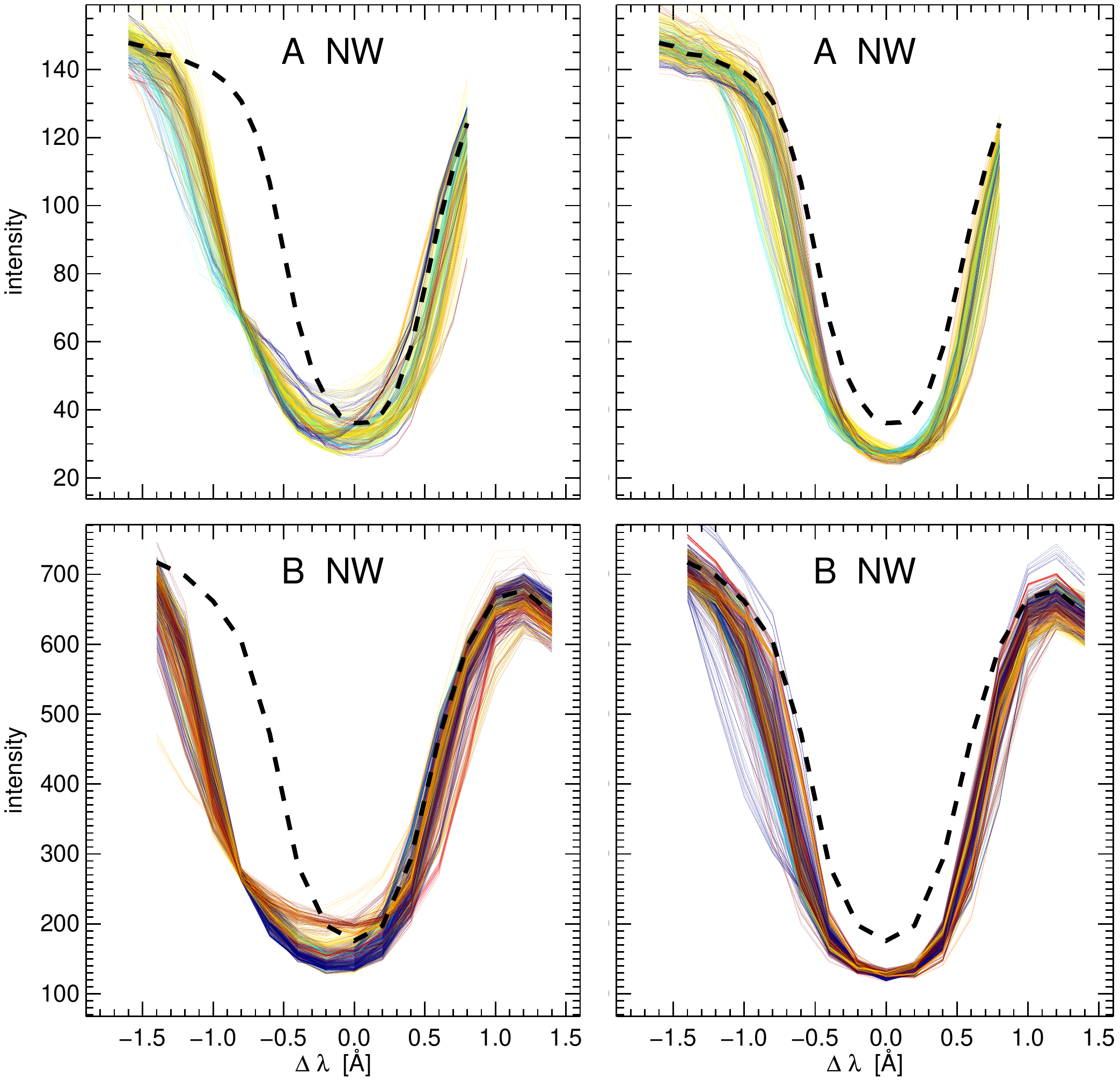}
  \caption[]{\label{fig:profiles} 
  \Halpha\ profiles for all pixels selected with the small green boxes
  in the scatter diagrams of Fig.~\ref{fig:darkdark} \reva{for data~A
  ({\em upper row}) and data~B ({\em lower row\/})}.
  \reva{{\em Dashed black curve\/}:} average profile over the field of
  view and the sequence duration.
  {\em First column\/}: profiles at the time of each darkest blue-wing
  sampling.  
  {\em Second column\/}: profiles at the time of each darkest core
  sampling.  
  \reva{{\em Color coding\/}: time delay between these samples on the
  scale of Fig.~\ref{fig:darkdark}.}
  The intensity units are the same as for the scatter diagrams in
  Fig.~\ref{fig:darkdark}.
  \reva{data~A:} 402 pixels with \reva{60\% within the yellow-orange}
  $2-5$~min \reva{positive-delay}  range. 
  \reva{data~B:} 635 pixels with \reva{26\% within} $2-5$~min delay.
  }
\end{figure}

\begin{figure}
  \centering
  \includegraphics[width=88mm]{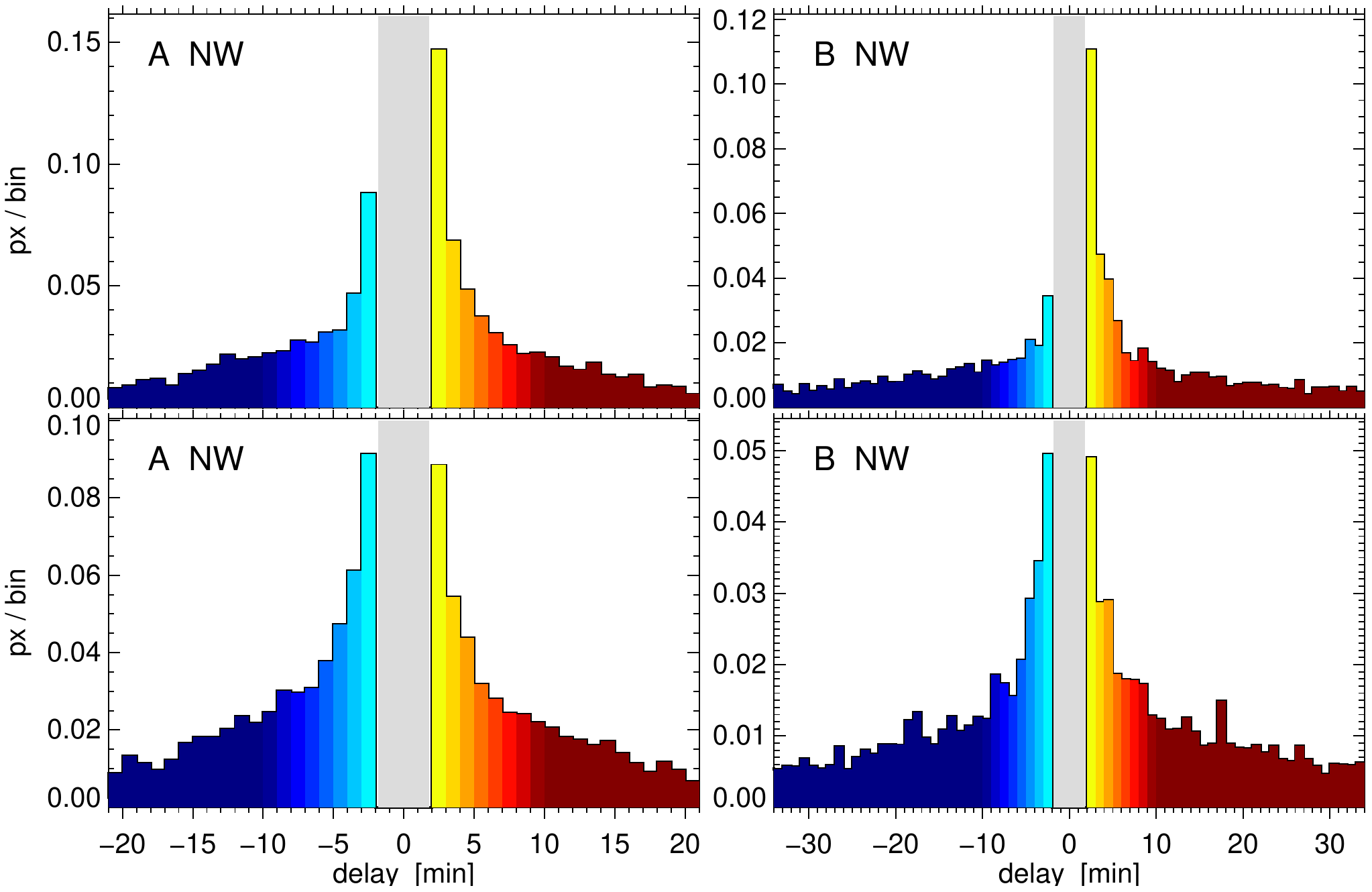}
  \caption[]{\label{fig:delays} 
  \reva{Histograms of sampling delays for data~A ({\em left\/}) and
  data~B ({\em right\/}). 
  The latter cover a twice longer delay range than shown here.
  {\em Upper row\/}: delays corresponding to} the image pairs in
  Fig.~\ref{fig:darkdark}, positive for darkest line-core
  \reva{occurrence} after darkest blue-wing \reva{occurrence} per
  network pixel, \reva{and negative for darkest core before darkest wing
  occurrence}. 
  {\em Lower row\/}: \revapar delays between the darkest blue-wing
  \reva{occurrence} and the second-darkest blue-wing \reva{occurrence}
  per network pixel.
  \reva{The delays are color-coded as in Fig.~\ref{fig:darkdark}.
  The gray rectangles mark the $[-2,+2]$~min ranges without delay
  determinations.}
  The bin widths are 1 min. 
  The $y$-axes \reva{specify numbers of pixels per bin as fraction of
  the network area.} 
  }
\end{figure}

RBE and RRE pixels are mutually exclusive, so that when type II
spicules cause subsequent core darkening, they contribute dark-dark
correlation at the left when they appear as RBE and at right when they
appear as RRE, \reva{with} each type \reva{contributing}
no-correlation samples for the opposite diagnostic.
For these samples the green contours should show no spatial overlap,
\reva{which can be checked with the second-row panel blinker in the
online material}.

Another contribution to the green selections comes from dynamic
fibrils that are treated further in Sect~\ref{sec:interpretation}. 
They occur mostly in the top third of the field where some contours
show overlap between columns. 
There are more non-overlap type II spicule contours in the lower part
of the field, but they generally occupy the same parts of the field of
view, which suggests that RBE and RRE launch along similar trajectories from
common origins while they upset the dark-dark correlation for each other.
The RRE discovery publication of
\citetads{2013ApJ...769...44S} 
indeed reported that many RREs are seen closely parallel to or
touching RBEs or transiting from one to the other.

The scatter triangle at the right reaches less deeply than the one at
the left, implying that the darkest RREs are less dark than the
darkest RBEs while \revc{they} still have as dark subsequent red-core
fibrils.  
This is also seen in the two wing images.
The tail of the distribution curve along the right-hand axis also
drops more steeply for RREs, which confirms that there are fewer RREs
than RBEs, as reported by \citetads{2013ApJ...769...44S}. 

\paragraph{Figures~\ref{fig:darkdark} -- \ref{fig:delays}.}
Next we ask a broader question: how does the overall \Halpha\ core
darkness relate to RBEs?
So far, we correlated RBEs and core redshifts per pixel at specified
delays of a few minutes, inspired by the \PubB\ PHE and its aftermath.
Figure~\ref{fig:darkdark} instead compares darkest outer-wing
instances \reva{per pixel} with darkest core instances at any delay
\reva{beyond two minutes, also negative. 
\revapar This inventory was made only for network} in view of
Fig.~\ref{fig:masked}.
Using the very darkest value sequence-wide implies selecting moments
of good seeing for both the wing and core samples.
We performed this for both data~A and B because they differ appreciably in
duration, seeing quality, and network extent and activity.

The leftmost images in Fig.~\ref{fig:darkdark} are again constructed
by assigning the darkest value to each pixel that it reached in the blue
wing during the entire sequence, in the same way as in Fig.~\ref{fig:darkest}, but only
for network areas.
The second-column images are similar constructs by selecting the
darkest value per pixel in the \revapar \Halpha\ core, \reva{using the
mean intensity} over $\Delta \lambda \tis -0.4$ to $+0.4$~\AA\ to
reduce the Doppler sensitivity as in a spectroheliogram with 0.8~\AA\
bandpass.
\reva{We excluded delays within the $[-2,+2]$~min range
because Fig.~\ref{fig:profiles} shows that many RBEs have dark cores
that would contribute same-feature sampling with self-correlation.
We avoided this contribution by not searching for the darkest core
occurrence within two minutes before or after each pixel's darkest
wing occurrence; this is longer than the RBE lifetimes.}

The darkest-core images also show fibrilar morphology, with remarkable
correspondence with the blue-wing images not only in overall
morphology, but in many places also in detail in darker features,
\reva{as is best seen with the panel blinkers in the online material}.
This is quantified by the \reva{significant} dark-dark
\reva{associations} in the scatter diagrams at the right. 
Thus, selection of the darkest instant of a blue-wing pixel gives a
good chance of its being part of a dark fibrilar feature made up by
nearby darkest samplings, possibly at other times, and also of being
part of a similar dark fibrilar feature in the line core made up by
darkest samples each at some other time than its darkest-wing
sampling.
\revapar

The \reva{significant} dark-dark \reva{association} and the image
similarities in Fig.~\ref{fig:darkdark} suggest that darkest RBEs and
darkest core fibrils have considerable commonality while excluding
simultaneity. 
We detail this further by analyzing the actual delays between each
pair of samplings per pixel.
They are shown in the color-coded maps in the third column of
Fig.~\ref{fig:darkdark}.
The many yellow-orange pixels for data~A (26\% of all network pixels)
marking $2-5$~min delays imply that a quarter of the darkest RBE
instances were followed by the darkest line-core instances within a
few minutes. This suggests direct causal relationship.
For data~B this filling factor is smaller (19\%), implying a higher
probability that the darkest core instance occurred earlier or later
and was not a direct result of the sampled RBE.
The duration of the data~B sequence was three times longer and its field more
active in producing RBEs, \reva{RREs, and also} the dynamic fibrils
discussed in Sect.~\ref{sec:discussion}, so that the chance that some
other darkest-core pixel would win from the darkest within $2-5$~min
delay is higher.

The yellow-orange pixels have contributed to the dark-dark
\reva{associations} of the previous figures, with the additional property that for
each diagnostic, its darkening reached the very darkest value for that
pixel during the whole sequence. 
They not only chart where dark fibrils follow on RBEs, but map the
darkest of both. 
Their aggregations also show much fibrilar morphology, with good
correspondence with the dark features defined by darkest samplings in
the images to the left. 

The \PubB\ contrail fibril shows a yellow-to-red retraction signature;
more such fibrilar yellow-red alignments can be seen elsewhere, but
with much confusion.

Figure~\ref{fig:profiles} shows the \Halpha\ profiles at both sample
moments for the very darkest-darkest pixels selected by the small
green boxes in the scatter diagrams \reva{in Fig.~\ref{fig:darkdark}}. 
In the \reva{left-hand} panels all curves cross close together for
$\Delta \lambda = -0.8$~\AA\ because the width of the selection
\reva{boxes is set} small to limit the number of profiles.
The \reva{right-hand panels do not show such crossing points} because
the averaging over core wavelengths accepts different profile shapes.
All these darkest-darkest profiles similarly represent RBE blue-wing
darkening from combined blueshift and widening at the left and more
symmetric dark-core signature at the right.  

\reva{The yellow-orange profiles with positive $[2-5]$~min delay (241
for data~A, 162 for data~B) are mostly similar. 
The blue negative-delay profiles show more blue-wing spread, light
blue in the first panel, and dark blue in the last panel. 
The latter may include some RBEs in this darkest-core selection because
some RBEs also have very dark cores.}

\reva{Figure~\ref{fig:delays} analyzes the delays in
Fig.~\ref{fig:darkdark} in terms of occurrence histograms in its
upper-row graphs.
They are similar for Data A and B. 
The histograms show systematic patterns that differ strongly from a
horizontal distribution for no association.
Both panels show a peak to the right of the gap that implies a high
probability that the core reached its darkest value a few minutes
after the darkest RBE.
For data~B, the peak reaches slightly lower filling fraction because the sequence lasted longer and there were more competing features.
Summing the $2-5$~min positive-delay bins gives the 26\% and 19\%
yellow-orange fill fractions in the delay maps in
Fig.~\ref{fig:darkdark}. 

The peaks left of the gaps correspond to the light-blue pixels in
Fig.~\ref{fig:darkdark} and are still considerable compared to the
outer tails.
For these pixels the darkest RBE instance was not followed by a
darkest core instance within a few minutes, but they had their darkest
core moment already just before the darkest RBE.
}

Farther away from the gaps, \reva{beyond six minutes,} the decays
reach symmetry between positive and negative \reva{values}. \revapar
\reva{They} imply that \reva{outside the $[-2,+2]$~min range the
core tends to be darkest preferentially at} shorter time difference
\reva{from} the darkest wing sampling, and that beyond \revc{six-minute}
difference the latter may have occurred later or earlier with equal
\reva{diminishing} probability.
This suggests recurrence, \revapar \reva{with the darkest} instances
occurring \reva{close after each other} in temporal groups.

The \reva{histograms in the second row of} Fig.~\ref{fig:delays}
support this suggestion.
For these the second darkest sampling is not one of the line core, but
again of the blue wing at $\Delta \lambda = -0.8$~\AA, \reva{also
beyond the $[-2,+2]$~min exclusion period}. 
\revapar \reva{Its} delay is a measure of RBE repetitions and the time
\reva{interval} between their darkest instances. 
\reva{Both histograms} show symmetric peaks adjacent to the gaps
\reva{and steep decays beyond, suggesting that darkest RBE
instances repeat close in time.}
\revapar Beyond $\pm 7$~min, the decay tails resemble those in the
upper \reva{panels}.

Thus, the very darkest RBEs tend to appear in close succession, and
have a good chance to produce maximum core darkening singly or in
tandem.
The repeat time indicated by the peaks in the lower
\reva{histogram} is \reva{$3-5$}~min.
\reva{This is longer than the mean} RBE recurrence time of of 84~s
reported by \citetads{2013ApJ...764..164S} 
\revc{because} $[-2,+2]$~min simultaneity is excluded.
With repeats faster than the typical persistence duration of
subsequent core darkenings (Figs.~\ref{fig:avertracks} and
\ref{fig:onions}), it may well be that darkest-core instances result
from summing aftermaths of multiple preceding type II spicules. 
\reva{Cases $e$ -- $g$ in Fig.~\ref{fig:selecttracks} seem examples
of such a buildup.}

\reva{Together, these histograms show that the darkest line-core
instance of any network pixel most probably followed within a few
minutes after the darkest RBE instance of that pixel. 
The next highest probability is that it followed on a previous
next darkest RBE instance a few minutes earlier.
When we add these second-best pairs by summing over the $[-5,-2]$ and
$[+2,+5]$~min delay ranges, the filling fractions for darkest cores
that are directly, likely causally, associated with darkest RBEs
increase to 43\% of all network pixels for data~A and to 27\% for
data~B.
}

\begin{figure}
  \centering
  \includegraphics[width=88mm]{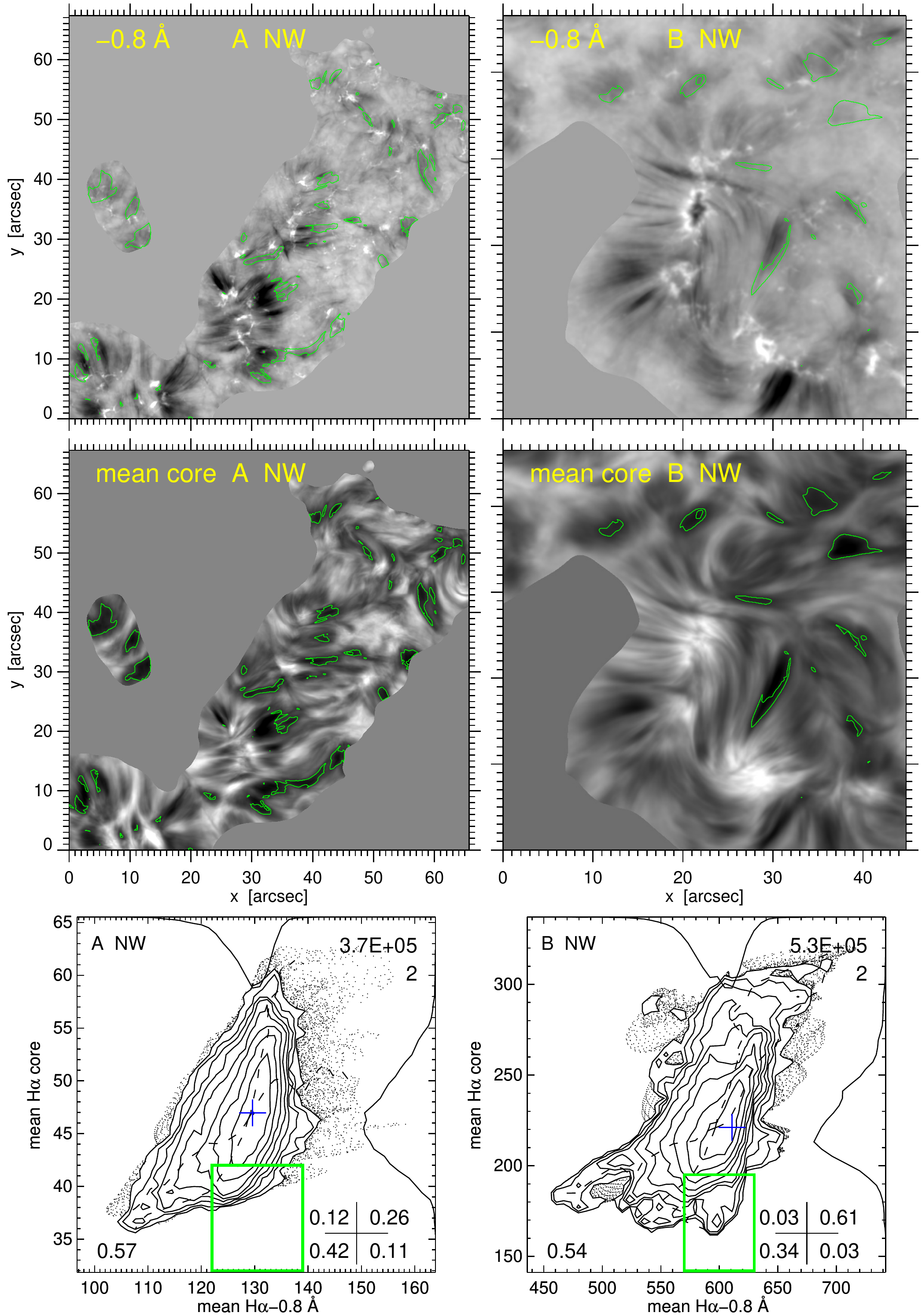}
  \caption[]{\label{fig:averall} 
  Temporal averages and scatter analysis for the network areas
  \reva{in data~A ({\em left:\/}) and data~B ({\em
  right\/})}. \revapar {\em First row\/}: mean images at
  $\Delta \lambda \tis -0.8$~\AA.
  {\em Second row\/}: mean images summing the \Halpha\ core over
  $\Delta \lambda \tis [-0.4,+0.4]$~\AA.
  {\em Third row\/}: corresponding scatter diagrams.
  The poor-seeing moments are discarded, as specified in
  Fig.~\ref{fig:seeing}.
  The green boxes select pixels with average wing darkening but large
  core darkening.
  \revb{Online image blinkers:
  \online{figure15_blink13.pdf}{1-3}, 
  \online{figure15_blink24.pdf}{2-4}}.
  }
\end{figure}

\begin{figure*}
  \centering
  \includegraphics[width=\textwidth]{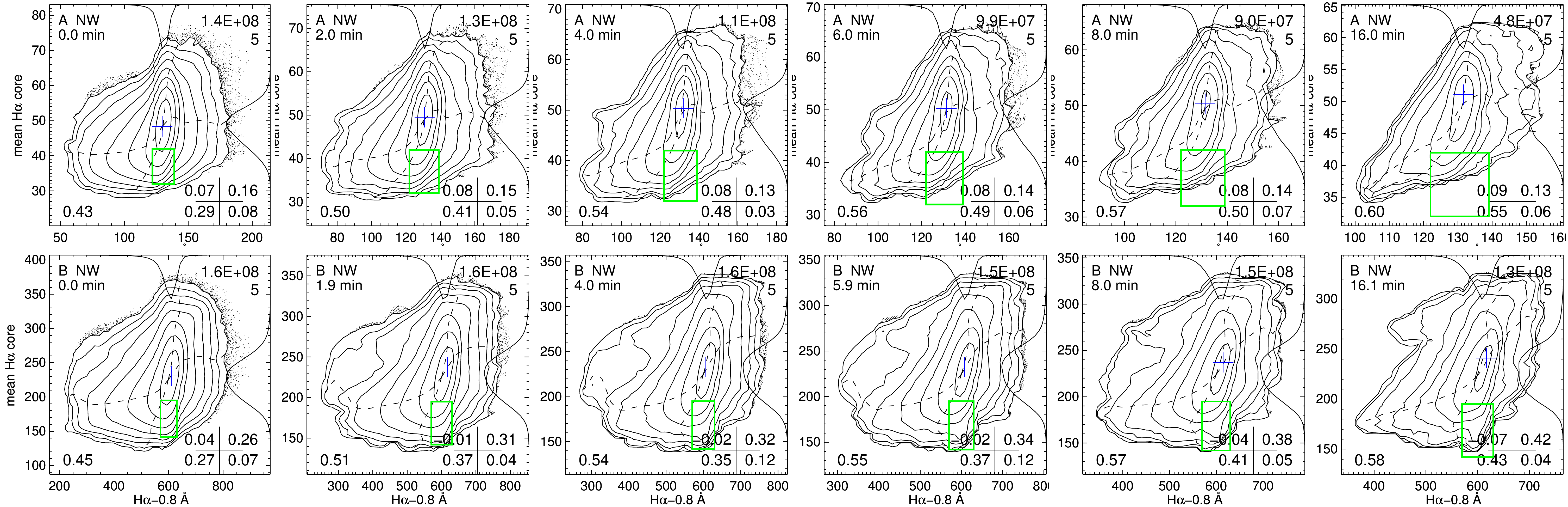}
  \caption[]{\label{fig:boxcar} 
  \reva{Full-sequence} scatter analysis \reva{for data~A ({\em upper
  row\/}) and data~B {\em lower row\/}). \revapar
  Format as} in Fig.~\ref{fig:onions}, but only for the network areas
  and with temporal boxcar averaging over the duration specified
  \reva{at the top left} in each panel.  
  The poor-seeing moments are discarded, as specified in
  Fig.~\ref{fig:seeing}.
  \revapar {\em Abscissas\/}: \Halpha\ wing intensity at
  $\Delta \lambda \tis -0.8$~\AA.
  {\em Ordinates\/}: \Halpha\ core intensity averaged over
  $\Delta \lambda \tis [-0.4,+0.4]$~\AA.
  The axes shrink along rows because momentary extremes \revc{lose} 
  weight. 
  \reva{The green boxes outline the same ranges as in Fig.~\ref{fig:averall}.}
  }
\end{figure*}

\paragraph{Figure~\ref{fig:averall}} \hspace*{-1.5ex}
\revc{compares} \reva{temporal} means over the entire image sequences except
for poor-seeing moments. 

\reva{In this averaging specific pixel-intensity combinations build up
signature when they are persistent or repeat in place, not when they
appear briefly only once or appear at a random location per image.
The short-lived \PubB\ PHE, contrail, and aftermath darkenings that are
so conspicuous in Fig.~\ref{fig:overview} do not leave much signature
in this temporal averaging.
Because the pore is the only long-lived dark feature in these data, the
other} dark streaks \reva{and blobs} in these time-averaged images
\reva{most probably} represent recurrent features.
\reva{The many darkest-wing darkest-core associations found in
Figs.~\ref{fig:darkdark} and \ref{fig:delays} also add up only when
they repeatedly occur in place, with RBEs occurring recurrently and
sequentially boosting line-core darkness}.
\revapar

The diagram for data~A at the left shows a tilt of the higher contours
from the summit pointing to and ending in the green box, whereas the
darkest-darkest contour spur reaches farther left. 
Its tip \revapar \reva{shows} the tightest darkest-darkest
\reva{association} of all scatter diagrams so far \reva{(note the
convergence of the moment curves)}, describing locations where darkest
time-averaged RBE presence goes together with darkest time-averaged
core fibril presence.  
\reva{The corresponding image blinker in the online material confirms
the high degree of correspondence between the darkest features.}

The green box instead selects pixels that sample dark core averages with
only slight or no averaged outer-wing darkening. 
The corresponding contours in the \reva{data~A} images above suggest
that many of these lie at the tips of RBEs. 
They may \reva{statistically} mark sites where aftermath core
darkening started beyond the visible RBE extent, as was the case for
\PubB\ contrail B, of which the longer post-RBE contrail extent is
evident \revapar in Fig.~\ref{fig:overview}.

\reva{In data~B the similarly selected pixels} come partly from the
large and darkest summed-core streak \reva{below and to the right of
the pore at image center. 
It} is incompletely covered by the corresponding summed-RBE feature,
similarly to the non-covered tips at left. \reva{However,} other
contributions come from dark core blobs in the upper third of the
network area that are without clear RBE counterparts, \reva{too round 
to be RREs, and are identified as} dynamic fibrils below. 
\reva{Together with RREs, they contribute to the conspicuous sag in
the outer contours contained in the selection box.}
\revapar

The higher contours of the scatter diagrams agree between the two data
sets. 
They show considerable dark-dark \reva{association} up to the mountain
summits and also considerable bright-bright \reva{association} above
them. 
The corresponding bright patches lie above network, whereas the
darkest RBE and core features lie around that and point away from it. 

\paragraph{Figure~\ref{fig:boxcar}.}
The final figure shows how the sequence-average scatter correlations
in Fig.~\ref{fig:averall} build up over time.  
\reva{For reference, all diagrams have green boxes with the same
pixel-pair value ranges as in Fig.~\ref{fig:averall}.}

The construction and format resemble Fig.~\ref{fig:onions}, but here
the core intensity is sampled as in a spectroheliogram, as in
Figs.~\ref{fig:darkdark} and \ref{fig:averall}, only the network areas
contribute, and instead of imposing temporal delays between the wing
and core samples, temporal boxcar averaging is applied to both over
increasing duration. 
The \reva{pixel-pair numbers are much higher, which produces smoother
contours} than for Fig.~\ref{fig:averall} because per panel, each
instant contributes for which the boxcar duration around it fits
within the sequence duration.
\reva{Features with specific pixel-value combinations that are
long lived or appear recurrently in place so gain much weight.
As in Fig.~\ref{fig:strousdemo}, the data~B pore is the only long-lived
dark feature 
in our data; SHOWEX inspection shows that it produced the slowly
developing left-pointing spur that becomes a promontory below the
boxcar duration label in the last panel.

The diagrams in the first column (no temporal averaging) are closely
similar to the second-column panels in Fig.~\ref{fig:onions},
they differ only in their core wavelength sampling and in the
full-field/network difference.} 
The 0 min panels in Fig.~\ref{fig:boxcar} show larger dark-dark
\reva{association} from sampling the mean core, which \reva{can be dark
in RBEs} \revapar (first column of Fig.~\ref{fig:profiles}). 
This same-instant \reva{association} contributes to the dark-dark
\reva{correspondences} in Fig.~\ref{fig:averall}, but was excluded in
Figs.~\ref{fig:darkdark} and \ref{fig:delays}.  

\reva{Along the rows in Fig.~\ref{fig:boxcar}, the scatter mountains
gradually transform from resembling the 0 min mountains in
Fig.~\ref{fig:onions} toward resembling those in
Fig.~\ref{fig:averall}.
With} increasing boxcar duration, the \reva{darkest-darkest
associations} tighten.
The findings in Figs.~\ref{fig:darkdark} and \ref{fig:delays} suggest
that \reva{this association is dominated} at 4 and 6 min boxcar
averaging \revapar by dark core fibrils directly following on RBEs and
that the \reva{association} tightening at longer averaging is
contributed by RBE\reva{-plus-core-darkening recurrence} at the same
locations.

\reva{The slowly developing contour bulge in the green boxes in the
lower row is attributed to dynamic fibrils in the next section. 
data~A in the upper row had fewer of these and shows the tightest
dark-dark mountain spur of all our scatter diagrams in the rightmost
diagram, quantified by the lower-left quadrant Pearson coefficient 0.55. 
It resembles the opposite bright-bright extension in the last panel of
Fig.~\ref{fig:strousdemo} in displaying a high degree of association.
For example, all RBE pixels darker than 120 data units have
darker-than-average cores in these 16 min boxcar summations, with high
darkest-darkest correspondence.
The corresponding contours extend fairly symmetrically along the
full-correlation diagonal so that the darkest pixels in the one are
also darkest in the other at similar densities.}

\section{Interpretation}\label{sec:interpretation}

\paragraph{Confusion from close recurrence.}
Figure~\ref{fig:selecttracks} and its online movie versions
illustrate the great difficulty in direct one-to-one recognition of RBE
outer-wing darkening followed by core darkening as in the \PubB\
PHE--contrail--aftermath pattern, but occurring more ubiquitously and
at smaller scale.
We attribute this difficulty not only to the very small and
seeing-sensitive scales, but also to the dynamical confusion caused by
frequent type II spicule recurrence, both cospatial and nearby, in the
\Halpha\ scenes around network. 
The RBE and RRE occurrence charts in
\citetads{2009ApJ...705..272R}, 
\citetads{2012ApJ...752..108S} ,
and \citetads{2013ApJ...769...44S}, 
the finding of fast recurrence in
\citetads{2013ApJ...764..164S} ,
and the $x-t$ time-line charts in
\citetads{2013ApJ...767...17Y} 
indeed suggest frequent repetitions in which new RBEs and RREs closely
follow tracks of previous ones or similar ones nearby.  
Moreover, the \PubB\ event had an adjacent similar feature (contrail B in
\PubB) as well as a subsequent one (contrail C in \PubB). 
The initial dark blobs in Fig.~\ref{fig:avertracks}, the delay
\reva{histograms} in Fig.~\ref{fig:delays}, and the time-averaged
RBE and fibrilar darkening in Fig.~\ref{fig:averall} and their
\revapar build-up in Fig.~\ref{fig:boxcar} also suggest
\reva{frequent} recurrence.
\revapar

\paragraph{Dynamic fibrils.}
These were first described by
\citetads{2006ApJ...647L..73H} 
and \citetads{2007ApJ...655..624D} 
and also occurred in our observations, especially in data~B.
\reva{They show up dark at line center but darken the outer wings only
slightly, so that they} do not contribute dark-dark correlation to
\reva{our} time-delay scatter diagrams. 
\revapar Inspection per SHOWEX showed that the roundish dark blobs in
the upper third part of the fourth image of Fig.~\ref{fig:averall},
some with green-contoured centers, correspond to patches of recurrent
dynamic fibrils, \reva{for example, the blob harboring the leftmost
contour and the blob-with-contour to its right, both below the
label}. 
Persistent repetitions at the same location build up \reva{their}
time-average core darkness in this image, with only a slight summed
outer-wing darkening in the second image of Fig.~\ref{fig:averall}.  

Dynamic fibrils also contributed green contours in
Fig.~\ref{fig:reversed} by being dark at line center without prior
darkening in either wing. 
These contours are likely to overlap, in contrast to RBEs versus RREs,
as indeed seen in the upper third of the first-row images in
Fig.~\ref{fig:reversed}.

With this identification, these dynamic-fibril patches can also be
recognized as showing a mottled short-streak morphology in the third
image of Fig.~\ref{fig:darkest} and the companion first lower-row
image of Fig.~\ref{fig:darkdark}. 
This morphology implies that only short dynamic-fibril segments
\reva{become darkest pixels} over time. 

There were fewer dynamic fibrils in the field of data~A.
\revapar In the lower row of Fig.~\ref{fig:boxcar}, the greater \revapar
contribution \reva{by dynamic fibrils and RREs} is recognized as
distinct sags in the bottom contours \reva{within the green selection
box}; it builds up with integration duration through recurrence. 
\reva{Without this contribution and the pore, the final diagram would
show cleaner darkest-darkest association, as in the upper row.}

\paragraph{Incomplete overlaps.}
There are multiple reasons why our statistical correlations cannot
reach one-to-one purity.
The first is the mutual exclusion of RBEs and RREs that is illustrated in the
second row of Fig.~\ref{fig:reversed}, where many of their locations
appear bright in the other wing panel.
A second is the presence of dynamic fibrils, which also contribute to the
green pixel selections in Fig.~\ref{fig:reversed}.
A third is the lateral motion of type II spicule features that upsets
time-delayed correlations per pixel. 
\citetads{2012ApJ...752..108S} 
and \citetads{2013ApJ...764..164S} 
reported transverse speeds for RBEs of $5-10$~\kms\ that may give
sizable offsets between cause and time-delayed effect.

Furthermore, RBEs become conspicuous in the outer blue \Halpha\ wing
only at appreciable distance from their \revc{launching sites in the network},
whereas the return flows retract closer to these roots. 
Their corresponding blue-wing and red-core darkening therefore tend to
lie adjacent along the type II spicule track rather than cospatial per
pixel. 
This is demonstrated by the leftward extent, left of the plus,
of the dark clouds in the bottom row of Fig.~\ref{fig:avertracks}, and
also by the first and third good-seeing columns of
Fig.~\ref{fig:onions-samples} and the second and fourth columns of
Fig.~\ref{fig:darkest}, where many box-selected dark features extend
closer to network in the delayed line-core images.
In addition to these missing RBE feet in overlap correlations, there
are also missing RBE tips, as suggested by Fig.~\ref{fig:averall}.

\section{Discussion} \label{sec:discussion}

\paragraph{Do type II spicules cause long network fibrils?}
The statistical approach of Fig.~\ref{fig:avertracks} and
Figs.~\ref{fig:onions}--\ref{fig:reversed}
brings evidence of return flows and fibrilar darkening after RBEs in
the form of significant dark-dark \reva{associations} for
multiple-minute delays.
They are far from one-to-one, but this is to be expected from mutual
RBE and RRE exclusion, presence of dynamic fibrils, missing overlaps
of feet and tips along type II spicule tracks, and confusion from fast
recurrences.
The remaining dark-dark \reva{associations} in the time-delay diagrams in
Figs.~\ref{fig:onions-samples}--\ref{fig:darkdark} are highly
significant, showing up as prominent left-down-sloping mountain spurs
in stark contrast to the absence of significant correlations in the
simultaneous diagrams in Figs.~\ref{fig:onions} and \ref{fig:darkest}.
These \reva{association} spurs imply that there is a strong tendency for
\Halpha\ RBEs to produce subsequent dark \Halpha\ fibrils
intermittently.  

The tightest dark-dark \reva{associations} are in
Figs.~\ref{fig:averall}--\ref{fig:boxcar} and are tighter than those
in Fig.~\ref{fig:onions}.
They imply that locations that become darkest in time-averaged \Halpha\
blue-wing from sampling repeated RBEs also tend to be darkest in the
time-averaged \Halpha\ line core. 
Frequent RBE recurrence was established by
\citetads{2013ApJ...764..164S} 
and is also diagnosed in Figs.~\ref{fig:avertracks}, \ref{fig:darkdark}
and \ref{fig:delays}.

Whether all \Halpha\ fibrils result from small-scale dynamic heating
events such as those producing RBEs and RREs remains an open question,
but our evidence leads us to conclude that type II spicules do
represent a major agent in the production of dark fibrils around
network.

The disparate numbers of RBEs and dark fibrils around network in
Fig.~\ref{fig:overview} appear contradictory, but because the
half-minute RBE lifetimes
(\citeads{2013ApJ...764..164S}) 
are much shorter than the multi-minute aftermaths in
Figs.~\ref{fig:onions} and Fig.~\ref{fig:delays}, such contrail fibrils
should indeed be more numerous at any one time. 
The densities of RBEs and core fibrils in the images in
Figs.~\ref{fig:darkest}--\ref{fig:darkdark} and Fig.~\ref{fig:averall}
indeed agree much better.

\paragraph{\Halpha\ and \CaIR\ extinctions.}
\Halpha\ is the strongest solar line in the visible off-limb spectrum
and so gave the chromosphere its name. 
On the disk it appears much weaker than \CaIIK, only about as strong
as \CaIR, but the fibrilar scenes it shows are so extraordinary that
\Halpha\ is the principal diagnostic of the chromosphere. 
We attribute this to its extraordinary atomic properties.
First, hydrogen is superabundant. 
Second, the line has exceedingly high excitation energy (10.2~eV). 
Third, it sits on top of the strongest line in the solar spectrum,
\Lyalpha. 
Together, these factors define its extinction coefficient with high
temperature sensitivity, including very large nonequilibrium
sensitivity that the \CaII\ lines do not have.
We give a brief summary with key references here; more detail is given
in \PubB\ and Rutten
(\citeyearads{2016A&A...590A.124R}, 
\citeyearads{2017IAUS..327....1R}, 
\citeyearads{2017A&A...598A..89R}). 

The Einstein relation for the ratio of collisional excitation and
deexcitation rates in \Lyalpha\ imparts very steep Boltzmann
temperature sensitivity to the population of the \HI\ $n \tis 2$ level
that defines the \Halpha\ extinction coefficient
(\linkpubbpage{7}{Fig.~7} of \PubB) as well as its nonequilibrium
sensitivity.
In any feature that is sufficiently thick for \Lyalpha\ to reach
\revc{radiative} detailed balance (\Halpha\ fibrils certainly qualify because
they are already optically thick in \Halpha), \revc{collisional} detailed
balance in \Lyalpha\ 
and LTE extinction of \Halpha\ are reached fast (seconds) at high
temperature but slow (minutes) at low temperature
(\linkadspage{2002ApJ...572..626C}{8}{Fig.~7} of
\citeads{2002ApJ...572..626C}). 
The effect of the latter slowness is commonly called ``nonequilibrium
hydrogen ionization'' , but the culprit is the large \Lyalpha\ jump,
whereas the ionization proceeds from the $n \tis 2$ level in
instantaneous statistical equilibrium, in a loop governed by
photoionization in the Balmer continuum, and cascade recombination
including photon losses in the Balmer lines
(\linkadspage{2002ApJ...572..626C}{4}{Fig.~3} of
\citeads{2002ApJ...572..626C}). 
This loop imparts NLTE over- or underionization set by the difference
between the local temperature and the 5300~K radiation temperature of
the impinging Balmer continuum from \reva{the deep photosphere}, but
these NLTE departures are small with respect to the gigantic
$n \tis 2$ overpopulations from slow collisional settling in \Lyalpha\
in gas that cools after having been hot.

The best demonstration so far is the extension to 2D MHD simulation by
\citetads{2007A&A...473..625L} 
of the fundamental 1D hydrodynamics simulation of
\citetads{2002ApJ...572..626C}. 
The last panel of \linkadspage{2007A&A...473..625L}{4}{Fig.~1} of
\citetads{2007A&A...473..625L} 
\revb{(\online{2007A+A...473..625L_fig1.mov}{movie
version} in the online material)} shows 12~dex overpopulations of the
$n \tis 2$ level that define \Halpha\ extinction in the cool-down phases
of acoustic shocks in the simulated internetwork and 8~dex
overpopulations after passages of magnetically guided shocks in the
simulated network that produce dynamic fibrils shown in
\linkadspage{2007A&A...473..625L}{6}{Fig.~3} of
\citetads{2007A&A...473..625L}. 
The next to last panel of their
\linkadspage{2007A&A...473..625L}{4}{Fig.~1} shows 2-3~dex additional
overionization by the Balmer continuum for the cool-down phases.
Their \linkadspage{2007A&A...473..625L}{5}{Figure~2}
\revb{(unpublished online \online{2007A+A...473..625L_fig2.mov}{movie
version})} shows that hydrogen ionization reaches only 10\% in the
shocks, but this partial ionization already suffices to produce
\reva{8--12~dex} \Halpha\ overextinction \revapar in the post-shock
cooling phases.

Dynamic fibrils and RBEs both represent PHEs, but the latter reach
higher temperature and hydrogen ionization. 
This is obviously the case for tips of \revbpar 
type II spicules, which frequently reach full ionization \revapar
(\citeads{2011Sci...331...55D}; 
\citeads{2016ApJ...820..124H}), 
but it also holds along on-disk RBEs.
\reva{We show this by comparing} their appearance in \reva{\Halpha\
and} \CaIR\ \reva{assuming Saha-Boltzmann extinction.
In hot features this assumption holds for both lines; it was indeed
valid for \Halpha\ in the hot phases of the simulated internetwork
shocks and dynamic fibrils of
\citetads{2002ApJ...572..626C} 
and \citetads{2007A&A...473..625L}.} 

\linkpubbpage{7}{Figure~7} of \PubB\ shows that the Saha-Boltzmann
curves for the extinction of these lines cross over near 6400~K at a
few percent hydrogen ionization.  
At lower temperatures the higher extinction of \CaIR\ explains that in
this line RBEs start earlier and closer to their network roots than in
\Halpha\ (\linkadspage{2012ApJ...752..108S}{3}{Fig.~1} of
\citeads{2012ApJ...752..108S} 
and \linkadspage{2013ApJ...764..164S}{10}{Fig.~10} of
\citeads{2013ApJ...764..164S}). 

For higher temperature, \Halpha\ extinction increases steeply through
its Boltzmann sensitivity to reach a maximum near 8000~K and 80\%
hydrogen ionization, whereas \CaII\ ionization causes steeply
decreasing \CaIR\ extinction down to only 1\% of the maximum \Halpha\
value.
RBEs extend considerably farther in \Halpha\ than in \CaIR, as also
shown in \linkadspage{2012ApJ...752..108S}{3}{Fig.~1} of
\citetads{2012ApJ...752..108S} 
and \linkadspage{2013ApJ...764..164S}{10}{Fig.~10} of
\citetads{2013ApJ...764..164S}. 
This additional extent implies temperatures above 7000~K and hydrogen
ionization above 20\% in these outer RBE parts
(\linkpubbpage{7}{Fig.~7} of \PubB), \reva{below the near 100\% value
in} the \PubB\ PHE and contrail that were not visible at all in \CaIR\
and \reva{above the 10\% value in} dynamic fibrils that extend
equally far in the two lines
(\linkadspage{2016ApJ...817..124S}{5}{Fig.~3} of
\citeads{2016ApJ...817..124S}). 

In subsequent cooling of this hot gas, \Halpha\ then initially retains
its strong high-temperature extinction, which implies NLTE overopacities \revapar
\reva{similar to}
the high values in the wakes of magnetoacoustic
shocks found by \citetads{2007A&A...473..625L}. 
Return fibrils \revc{therefore} 
maintain \Halpha\ darkness even when the gas cools
below 6400~K and regains \CaIR\ presence, as was the case in the
\PubB\ aftermath at the very end of the redshifted retraction phase
(\linkpubbpage{6}{Fig.~5} of \PubB).

\paragraph{\Halpha\ and \Lyalpha\ scattering.}
The \Halpha\ source function, which together with the extinction
defines \Halpha\ image formation, is easier to describe because it
is dominated by resonance scattering, as for any strong chromospheric
line.   
\citetads{1957ApJ...125..260T} 
and
\citetads{1959ApJ...129..401J} 
described it as special in being ``photoelectrically controlled'' by
the Balmer ionization loop, but this contribution is only minor, while
resonance scattering represents the main agent
(\linkadspage{2012A&A...540A..86R}{6}{Sect.~6} of
\citeads{2012A&A...540A..86R}). 
The actual abnormality of \Halpha  \ is strong backscattering from
the chromosphere to the low photosphere over the very deep \Halpha\
opacity gap in the upper photosphere
(\citeads{1972SoPh...22..344S}), 
which produces a \Halpha\ source function raise there
(\linkadspage{2012A&A...540A..86R}{6}{Fig.~7} of
\citeads{2012A&A...540A..86R}) 
that was misinterpreted with the famous
\linkadspage{1959ApJ...129..401J}{5}{Fig.~3} of
\citetads{1959ApJ...129..401J}. 

The scattering nature of \Halpha   \ implies that features that are more opaque
become darker: \Halpha\ fibril darkness is defined by the local
extinction, not by the local temperature. 
The same holds for \CaIR,\ which scatters very similarly.
For \Halpha\ nonequilibrium extinction with retarded decay from
preceding large high-temperature values enhances fibril darkening.

Resonance scattering also increases the apparent size of observed
features.
The PHEs at the root of the RBE phenomenon are due to plasma processes
that possibly operate on physical scales below telescopic resolution
limits and heat gas along very thin tracks initially.
\Lyalpha\ surround-scattering into cool gas around these smears the
corresponding \Halpha\ opacity feature over tens of kilometers
(\linkadspage{2016A&A...590A.124R}{6}{Fig.~3} of
\citeads{2016A&A...590A.124R}) ,
while \Halpha\ scattering into cool surrounding gas smears its source
function signature over hundreds of kilometers.
These scattering diffusions together may contribute to RBE visibility
at the SST resolution.

The thin bright RBE stripes in the third column of
Fig.~\ref{fig:selecttracks} appear narrower because they mainly
represent core Dopplershifts, which are primarily contributed by the
last \Halpha\ photon scatterings on their way out and are widened
primarily by \Lyalpha\ surround-scattering, which smoothes their opacity. 
Their observed thinness suggests that the PHEs that underlie RBE
formation are intrinsically narrower than the observed blue-wing RBEs.

\paragraph{Inadequate resolution.}
The long reach of the mountain spur of dark-dark \reva{association} in
the scatter diagrams up to slightly dark -- slightly dark \reva{association}
close to the mountain summits is intriguing. 
Figure~\ref{fig:masked} was made to test whether these upper spur
parts come from quiet internetwork areas.
Such areas are not often seen in \Halpha\ images because most
internetwork is hidden under overlying canopies of \Halpha\ fibrils.
The only study addressing one at SST resolution so far is the one by
\citetads{2007ApJ...660L.169R} ,
who attributed the extremely dynamic very fine mottles they observed
with the SST to shocks. 
Internetwork shocks are well known and well understood in their
manifestation as \CaII\ \KtwoV\ and \HtwoV\ grains
(\citeads{1991SoPh..134...15R}; 
\citeads{1997ApJ...481..500C}), 
and because these shocks were also well demonstrated to act as
nonequilibrium \Halpha\ opacity enrichers by
\citetads{2002ApJ...572..626C} 
and \citetads{2007A&A...473..625L},
we expect to find a delay signature also from these in our scatter
diagrams.  

The internetwork scatter diagram at the left in Fig.~\ref{fig:masked} does
show some dark-dark \reva{association,} but far less so than the network
diagram at the right.
Figure~\ref{fig:masked} therefore indicates that also the
slightly dark -- slightly dark \reva{associations} along the upper
spur parts, including the first-moment crossing, come from the network
areas and are likely also due to magnetic processes.
This suggests that many or most of such magnetodynamical heating
features are not resolved even at the SST, resulting in summit-ward
migration along the spur, as shown for poor seeing in
Fig.~\ref{fig:onions-samples}. 
This lack of resolution is also suggested by the very fine striations
of the larger dark fibril bunches in Fig.~\ref{fig:selecttracks} and
its movie versions, which become indistinguishable at poor-seeing
moments.
\citetads{2012ApJ...752..108S} 
also suggested that their high-quality SST observations were not
resolving all actual RBEs. 

\paragraph{Intermittent hydrogen ionization.}
Our results, in particular the tight dark-dark \reva{associations} in
Figs.~\ref{fig:averall} and \ref{fig:boxcar}, are testimony of
intermittent heating including partial hydrogen ionization around our
quiet-Sun network areas. 
The RBE blue-wing darkenings directly portray such heating events.
Their \Halpha\ opacity stems from the steep increase of \Halpha\
extinction with temperature and hydrogen ionization.  
They are short lived but occur frequently and recurrently, and so build
up their time-averaged darkness. 
The \Halpha\ core darkenings at the same locations represent lower
temperatures but with high opacities, hence low intensities, that are
leftovers of preceding heating events. 
Without prior heating they would not have such a high \Halpha\ opacity.

Thus, the \Halpha\ chromosphere should be seen as the solar atmosphere
domain where hydrogen ionizes intermittently, doing so frequently and
repetitively on very small spatial and short timescales imposed by
magnetodynamical processes. 
In classical 1D static standard models such as ALC7 of
\citetads{2008ApJS..175..229A},
the chromosphere is a plane-parallel layer in which the degree of
hydrogen ionization increases gradually from fully neutral at the
bottom to fully ionized at the top, at infinite horizontal extent and
eternal constancy.
Our interpretation instead makes the network chromosphere
intrinsically 3D structured and time-dependent, intermittently heated,
and achieving hydrogen ionization by magnetic agents operating on very
small spatial and very short temporal scales.

\section{Conclusion}  \label{sec:conclusion}
We have collected evidence that long network-surrounding \Halpha\
fibrils often follow on the launch of propagating heating events that
are observed as rapid blue and red excursions and \revc{represent}
on-disk manifestations of type II spicules.
We suggest that partial hydrogen ionization is a principal agent in
defining the subsequent fibril visibility along the tracks of these
events.
The evidence further suggests that many of these heating events are
not yet resolved by the currently best solar telescope. 

Our suggestion is that the \Halpha\ chromosphere is an extremely
dynamic solar-atmosphere domain where hydrogen frequently ionizes at
least partially on very small scales due to magnetic dynamism  (even
in quiet-Sun areas)  and that this intermittent heating causes
the extraordinarily rich fibrilar appearance of the \Halpha\
chromosphere.
The moral is that in modeling \Halpha\ network fibrils we may no
longer assume statistical equilibrium, but instead must be concerned
with the local gas history, including what happened nearby in
\Lyalpha. 
This may also hold for other features that are seen particularly well in
\Halpha,\ such as Ellerman bombs, surges, filaments, and prominences,
and it may well hold for all chromosphere modeling because hydrogen
ionization is its dominant entity.

Progress may come from both observations and simulations.
The \acl{DKIST} promises higher resolution in \Halpha\ than the
\acl{SST}. 
When full-baseline solar imaging is developed at the \acl{ALMA}, this
may become the champion because the millimeter continua gain
nonequilibrium opacity in dynamic heating and cooling in the same
manner as \Halpha\ (\citeads{2017A&A...598A..89R}). 

In numerical simulations progress will come from including more
small-scale heating processes such as ion-neutral interactions
implemented in Bifrost by
\citetads{2018ApJ...860..116M} 
and also accounting for nonequilibrium hydrogen partitioning.
The subsequent \Halpha\ synthesis must be similarly time-dependent and
3D, which already poses a great challenge
(\citeads{2019AdSpR..63.1434P}), 
but additionally implement nonequilibrium \Halpha\ opacity
evaluation.

We look forward to such developments.

\begin{acknowledgements}
We thank Tiago Pereira for help during the data~B observations.
The SST is operated on the island of La Palma by the Institute
for Solar Physics of Stockholm University in the Spanish Observatorio
del Roque de los Muchachos of the Instituto de Astrof{\'\i}sica de
Canarias.
The Institute for Solar Physics is supported by a grant for research
infrastructures of national importance from the Swedish Research
Council (registration number 2017-00625). 
This research is supported by the Research Council of Norway, project
number 250810, and through its Centres of Excellence scheme, project
number 262622. 
We made much use of the SolarSoft and ADS libraries.
\end{acknowledgements}

\bibliographystyle{aa-note}
\bibliography{36113}

\end{document}

%% file: rrmacros-pub.tex
\definecolor{bllsht}{rgb}{0.40, 0.05, 0.01} 
\long\def\@ #1@{\par\vspace{0.5ex}\noindent
  {\Large \textcolor{red}{@~}}\color{bllsht}{#1}\color{black}\\[0.5ex]}
 
\makeatletter
\newcommand{\bibnote}[2]{\global\@namedef{#1note}{#2~}}
\newcommand{\biblink}[2]{\global\@namedef{#1link}{#2}}
\makeatother

\long\def\reva#1{{\color{red}{\bf #1}\color{black}}}  
\def\revapar{{\color{red}{\P\ }\color{black}}} 
\definecolor{amber}{rgb}{1.0, 0.49, 0.0}
\long\def\revb#1{{\color{amber}{\bf #1}\color{black}}}  
\def\revbpar{{\color{amber}{\P\ }\color{black}}} 
\long\def\revc#1{{\color{red}{\bf #1}\color{black}}}  

\bibpunct{(}{)}{;}{a}{}{,}    
\makeatletter
  \protected\def\sppresslink{\def\hyper@linkstart##1##2{}\let\hyper@linkend\@empty}
  \newcommandtwoopt{\citeads}[3][][]{%
   \href{http://ui.adsabs.harvard.edu/abs/#3/abstract}%
        {\sppresslink \citealp[#1][#2]{#3}}
   \biblink{#3}{\href{http://ui.adsabs.harvard.edu/abs/#3/abstract}{ADS}}}
 \newcommandtwoopt{\citepads}[3][][]{%
   \href{http://ui.adsabs.harvard.edu/abs/#3/abstract}%
        {\sppresslink \citep[#1][#2]{#3}}
   \biblink{#3}{\href{http://ui.adsabs.harvard.edu/abs/#3/abstract}{ADS}}}
 \newcommandtwoopt{\citetads}[3][][]{%
   \href{http://ui.adsabs.harvard.edu/abs/#3/abstract}%
        {\sppresslink \citet[#1][#2]{#3}}
  \biblink{#3}{\href{http://ui.adsabs.harvard.edu/abs/#3/abstract}{ADS}}}
 \newcommandtwoopt{\citeyearads}[3][][]{%
   \href{http://ui.adsabs.harvard.edu/abs/#3/abstract}%
        {\sppresslink \citeyear[#1][#2]{#3}}
   \biblink{#3}{\href{http://ui.adsabs.harvard.edu/abs/#3/abstract}{ADS}}}
\makeatother

\def\linkadspage#1#2#3{\href{http://adsabs.harvard.edu/cgi-bin/nph-data_query?bibcode=#1\&link_type=ARTICLE\&db_key=AST\#page=#2}{#3}}

\newcount\longrefs
\def\adv{\ifnum\longrefs=1 {Adv.\ Space Res.} \else 
                           {Adv.\ Sp'\ Res.}\fi}
\def\aap{\ifnum\longrefs=1 {Astron.\ Astrophys.}\else 
                           {A\hbox{\rm \&}A}\fi}
\def\aapr{\ifnum\longrefs=1 {Astron.\ Astrophys.\ Rev.}\else 
                            {A\hbox{\rm \&}AR}\fi}
\def\aaps{\ifnum\longrefs=1 {Astron.\ Astrophys.\ Suppl.}\else 
                            {A\hbox{\rm \&}A Suppl.}\fi}
\def\actaa{\ifnum\longrefs=1 {Acta Astronomica}\else
                            {Acta Astron.}\fi}
\def\aipcs{\ifnum\longrefs=1 {Am.\ Inst.\ Phys.\ Conf.\ Series}\else
                             {AIP Conf.\ Ser.}\fi}
\def\aj{\ifnum\longrefs=1 {Astron.\ J.}\else 
                          {AJ}\fi} 
\def\ao{\ifnum\longrefs=1 {Applied Optics}\else 
                           {Appl.\ Opt.}\fi} 
\def\aspcs{\ifnum\longrefs=1 {Astron.\ Soc.\ Pacific Conf.\ Series}\else 
                           {ASP Conf.\ Ser.}\fi} 
\def\apj{\ifnum\longrefs=1 {Astrophys.\ J.}\else 
                           {ApJ}\fi} 
\def\apjl{\ifnum\longrefs=1 {Astrophys.\ J. Lett.}\else 
                            {ApJL}\fi} 
\def\aplett{\ifnum\longrefs=1 {Astrophys.\ J. Lett.}\else 
                            {ApJ}\fi} 
\def\apjs{\ifnum\longrefs=1 {Astrophys.\ J. Suppl.}\else 
                            {ApJS}\fi}
\def\apss{\ifnum\longrefs=1 {Astrophys.\ Space Sci.}\else 
                            {Astrophys.\ Space Sci.}\fi}
\def\araa{\ifnum\longrefs=1 {Ann.\ Rev.\ Astron.\ Astrophys.}\else 
                            {ARA\hbox{\rm \&}A}\fi}
\def\azh{\ifnum\longrefs=1 {Astronomicheskii Zhurnal}\else 
                            {Astron.\ Zhur.}\fi}
\def\baas{\ifnum\longrefs=1 {Bull.\ Am.\ Astron.\ Soc.}\else 
                            {BAAS}\fi}
\def\bain{\ifnum\longrefs=1 {Bull.\ Astronom.\ Institutes Netherlands}\else
                            {Bull.\ Astr.\ Inst.\ Neth.}\fi}
\def\cjaa{\ifnum\longrefs=1 {Chin.\ J.\ Astron.\ Astrophys.}\else 
                            {Chin.\ J.\ A\&A}\fi}
\def\gca{\ifnum\longrefs=1 {Geochim.\ Cosmochim.\ Acta}\else 
                           {Geochim.\ Cosmochim.\ Acta}\fi}
\def\grl{\ifnum\longrefs=1 {Geophys.\ Res.\ Lett.}\else 
                           {Geoph.\ Res.\ Lett.}\fi}
\def\iaucirc{\ifnum\longrefs=1 {IAU Circulars}\else 
                          {IAU Circ.}\fi}
\def\icarus{\ifnum\longrefs=1 {Icarus}\else 
                          {Icarus}\fi}
\def\ip{\ifnum\longrefs=1 {in press}\else 
                          {in press}\fi}
\def\jcap{\ifnum\longrefs=1 {Jour.\ Cosmology Astropart.\ Phys.}\else 
                          {JCAP}\fi}
\def\jgr{\ifnum\longrefs=1 {J.\ Geophys.\ Res.}\else 
                           {J.\ Geophys.\ Res.}\fi}  
\def\jrasc{\ifnum\longrefs=1 {J.\ Royal Astron.\ Soc.\ Canada}\else 
                             {JRAS Can.}\fi}  
\def\memsai{\ifnum\longrefs=1 {Mem.~Soc.~Astron.~Italiana}\else
                              {MmSAI}\fi}
\def\mnras{\ifnum\longrefs=1 {Mon.\ Not.\ Roy.\ Astron.\ Soc.}\else 
                             {MNRAS}\fi} 
\def\na{\ifnum\longrefs=1 {New Astronomy}\else 
                          {New Astron.}\fi}
\def\nar{\ifnum\longrefs=1 {New Astronomy rev.}\else 
                           {New Astron.\ Rev.}\fi}
\def\nat{\ifnum\longrefs=1 {Nature}\else 
                           {Nat}\fi}
\def\pasa{\ifnum\longrefs=1 {Pub.\ Astron.\ Soc.\ Australia}\else 
                            {PASA}\fi} 
\def\pasj{\ifnum\longrefs=1 {Pub.\ Astron.\ Soc.\ Japan}\else 
                            {PASJ}\fi} 
\def\pasp{\ifnum\longrefs=1 {Pub.\ Astron.\ Soc.\ Pacific}\else 
                            {PASP}\fi} 
\def\physscr{\ifnum\longrefs=1 {Physica Scripta}\else 
                               {Phys.\ Scrip.}\fi} 
\def\planss{\ifnum\longrefs=1 {Planetary \& Space Science}\else 
                              {Plan. \& Space Sci.}\fi} 
\def\pre{\ifnum\longrefs=1 {Phys.\ Rev.\ E}\else
                           {Phys.\ Rev.\ E}\fi}
\def\procspie{\ifnum\longrefs=1 {Proc.\ SPIE}\else 
                                {Proc.\ SPIE}\fi} 
\def\qjras{\ifnum\longrefs=1 {Quarterly J.\ Royal Astron.\ Soc.}\else 
                             {QJRAS}\fi} 
\def\rmxaa{\ifnum\longrefs=1 {Revista Mexicana de Astron.\ y Astrofys.}\else 
                             {RMxAA}\fi} 
\def\sa{\ifnum\longrefs=1 {Soviet Astron..}\else 
                          {Sov.\ Astron.}\fi}
\def\skytel{\ifnum\longrefs=1 {Sky \& Telescope}\else 
                              {Sky \& Tel.}\fi} 
\def\solphys{\ifnum\longrefs=1 {Solar Phys.}\else 
                               {SoPh}\fi}
\def\sovast{\ifnum\longrefs=1 {Soviet Astron.}\else 
                              {Sov.\ Ast.}\fi}
\def\ssr{\ifnum\longrefs=1 {Space Sci.\ Rev.}\else 
                           {Space Sci.\ Rev.}\fi}
\def\zap{\ifnum\longrefs=1 {Zeitschr.\ f.\ Astrophysik}\else
                               {Z.\ Astrophys.}\fi}

\def\wlRRtop#1{\href{http://www.staff.science.uu.nl/~rutte101}{#1}}

\def\nl{,\ } 

\def\ITA{Institute of Theoretical Astrophysics\nl
         University of Oslo\nl
         P.O.\ Box 1029 Blindern\nl
         N-0315 Oslo\nl
         Norway}

\def\LA{Lingezicht Astrophysics\nl 't Oosteneind 9\nl 4158\,CA Deil\nl 
        The Netherlands}
\def\LMSAL{Lockheed-Martin Solar and Astrophysics Laboratory\nl
           3251 Hanover Street\nl Palo Alto, CA~94304\nl USA}

\def\RoCS{Rosseland Centre for Solar Physics\nl
          University of Oslo\nl
          P.O.\ Box 1029 Blindern\nl
          N-0315 Oslo\nl
          Norway}

\newacro{AA}{Astronomy \& Astrophysics}  
\newacro{ADS}{Astrophysics Data System}
\newacro{AIA}{Atmospheric Imaging Assembly}
\newacro{ALMA}{Atacama Large Millimeter/submillimeter Array}
\newacro{AO}{adaptive optics}
\newacro{ApJ}{Astrophysical Journal}
\newacro{AR}{active region}
\newacro{bb}{bound-bound}
\newacro{bf}{bound-free}
\newacro{BFI}{Broad-band Filter Imager}
\newacro{CE}{coronal equilibrium}
\newacro{CfA}{Center for Astrophysics}
\newacro{CME}{coronal mass ejection}
\newacro{CRD}{complete redistribution}
\newacro{CRISP}{CRisp Imaging SpectroPolarimeter}
\newacro{CRISPEX}{CRisp SPectral EXplorer}
\newacro{CS}{coherent scattering}
\newacro{DEM}{Differential Emission Measure}
\newacro{DF}{dynamic fibril}
\newacro{DKIST}{Daniel K. Inouye Solar Telescope}
\newacro{DLR}{Deutsches Zentrum f\"ur Luft- und Raumfahrt}
\newacro{DOT}{Dutch Open Telescope}
\newacro{DST}{Richard B. Dunn Solar Telescope}   
\newacro{EB}{Ellerman bomb}
\newacro{EDP}{\'{E}dition Diffusion Presse}  
\newacro{EIT}{Extreme ultraviolet Imaging Telescope}
\newacro{EPIC}{European participation in Solar-C}
\newacro{ERC}{European Research Council}
\newacro{ESA}{European Space Agency}
\newacro{EST}{European Solar Telescope}
\newacro{EUV}{extreme ultraviolet}
\newacro{FAF}{flaring active-region fibril}
\newacro{ff}{free-free}
\newacro{FITS}{Flexible Image Transport System}
\newacro{FOV}{field of view}
\newacro{fov}{field of view}
\newacro{FWHM}{full width at half maximum}
\newacro{HAO}{High Altitude Observatory}
\newacro{HD}{hydrodynamics}
\newacro{Hi-C}{High Resolution Coronal Imager Sounding Rocket}
\newacro{HMI}{Helioseismic and Magnetic Imager}
\newacro{IAA}{Instituto de Astrof\'{i}sica de Andaluc\'{i}a}
\newacro{IAC}{Instituto de Astrof\'{i}sica de Canarias}
\newacro{IAS}{Institut d'Astrophysique Spatiale}
\newacro{IAU}{International Astronomical Union}
\newacro{IBIS}{Interferometric Bi-dimensional Spectrometer}
\newacro{IDL}{Interactive Data Language}
\newacro{IMaX}{Imaging Magnetograph eXperiment}
\newacro{INAF}{Istituto Nazionale di Astrofisica}
\newacro{IB}{IRIS bomb}
\newacro{IR}{infrared}
\newacro{IRIS}{Interface Region Imaging Spectrograph}
\newacro{ISAS}{Institute of Space and Astronautical Science}
\newacro{ISP}{Institute for Solar Physics}
\newacro{ISS}{International Space Station}
\newacro{ISSI}{International Space Science Institute}
\newacro{ITA}{Institute for Theoretical Astrophysics}
\newacro{JAXA}{Japan Aerospace Exploration Agency}
\newacro{JSOC}{Joint Science Operations Center}
\newacro{KIS}{Kiepenheuer--Institut f\"{u}r Sonnenphysik}
\newacro{KPNO}{Kitt Peak National Observatory}
\newacro{LASP}{Laboratory for Atmospheric and Space Physics}
\newacro{LC}{liquid cristal}
\newacro{LMSAL}{Lockheed Martin Solar and Astrophysics Labratory}
\newacro{LOS}{line of sight}
\newacro{LTE}{local thermodynamic equilibrium}
\newacro{MC}{magnetic concentration}
\newacro{MCAO}{multi-conjugate adaptive optics} 
\newacro{MDI}{Michelson Doppler Imager}
\newacro{ME}{Milne-Eddington} 
\newacro{MHD}{magnetohydrodynamics}
\newacro{MOMFBD}{Multi-Object Multi-Frame Blind Deconvolution}
\newacro{MPE}{Max--Planck--Institut f\"ur extraterrestrische Physik}
\newacro{MPG}{Max--Planck--Gesellschaft}
\newacro{MPS}{Max Planck Institute for Solar System Research}
\newacro{MSSL}{Mullard Space Science Laboratory}
\newacro{MTF}{modulation transfer function}
\newacro{NAOJ}{National Astronomical Observatory of Japan}
\newacro{NASA}{National Aeronautics and Space Administration}
\newacro{NIST}{National Institute of Standards and Technology}
\newacro{NLTE}{non-local thermodynamic equilibrium}
\newacro{NLFFF}{non-linear force-free field}
\newacro{NOAA}{National Oceanic and Atmospheric Administration}
\newacro{non-E}{non-equilibrium}
\newacro{NSO}{National Solar Observatory}
\newacro{NWO}{Netherlands Organisation for Scientific Research}
\newacro{PHE}{propagating heating event}
\newacro{PRD}{partial redistribution}
\newacro{PROBA2}{PRoject for OnBoard Autonomy}
\newacro{PSBE}{post Saha-Boltzmann extinction}
\newacro{PSF}{point spread function}
\newacro{QS}{quiet Sun}
\newacro{QSEB}{quiet-Sun Ellerman-like brightening} 
\newacro{RAL}{Rutherford Appleton Laboratory}
\newacro{RBE}{rapid blue-shifted excursion}
\newacro{R-MHD}{radiation hydrodynamics}
\newacro{rms}{root mean square}
\newacro{RMS}{root mean square}
\newacro{ROB}{Royal Observatory of Belgium}
\newacro{ROI}{region of interest}
\newacro{RRE}{rapid red-shifted excursion}
\newacro{RTE}{radiative transfer equation}
\newacro{RTSA}{Radiative Transfer in Stellar Atmospheres}
\newacro{SCF}{slender \CaIIH\ fibril}
\newacro{SE}{statistical equilibrium}
\newacro{SB}{Saha Boltzmann}
\newacro{SDO}{Solar Dynamics Observatory}
\newacro{SJI}{slit-jaw image}
\newacro{SLI}{slit image}
\newacro{SNR}{signal-to-noise ratio}
\newacro{SO}{Solar Orbiter}
\newacro{SoHO}{Solar and Heliospheric Observatory}
\newacro{SP}{Spectropolarimeter}
\newacro{SST}{Swedish 1-m Solar Telescope}
\newacro{SUMER}{Solar Ultraviolet Measurements of Emitted Radiation}
\newacro{SUFI}{Sunrise Filter Imager}
\newacro{SVD}{singular value decomposition}
\newacro{SVST}{Swedish Vacuum Solar Telescope}
\newacro{STX}{Solar Telescope X}
\newacro{THEMIS}{T\'{e}lescope H\'{e}liographique pour l'Etude du 
   Magn\'{e}tisme et des Instabilit\'{e} Solaires}     
\newacro{TR}{transition region}
\newacro{TRACE}{Transition Region and Coronal Explorer}
\newacro{TSI}{total solar irradiance}
\newacro{UT}{Universal Time}
\newacro{UV}{ultraviolet}
\newacro{VAULT}{Very high angular resolution ultraviolet telescope}
\newacro{VIRGO}{Variability of solar IRradiance and Gravity Oscillations}
\newacro{VTT}{Vacuum Tower Telescope}    
\newacro{XRT}{X-Ray Telescope}

\long\def\startignore #1\stopignore{}   

\def\rmit#1{{\it #1}}              
\def\etal{\rmit{et al.}}           
           
\def\eg{\rmit{e.g.,}}              

\def\specchar#1{\uppercase{#1}}    
\def\specand{ and }                
\def\specand{\,\&\,}               

\def\CaII{\mbox{Ca\,\specchar{ii}}}

\def\HI{\mbox{H\,\specchar{i}}} 
 
\def\HeI{\mbox{He\,\specchar{i}}}

\def\Halpha{\mbox{H\hspace{0.1ex}$\alpha$}} 

\def\Lyalpha{\mbox{Ly$\hspace{0.2ex}\alpha$}}

\def\CaIIK{\mbox{Ca\,\specchar{ii}\,\,K}}       
\def\CaIIH{\mbox{Ca\,\specchar{ii}\,\,H}}

\def\HK{\mbox{H{\specand}K}}

\def\KtwoV{\mbox{K$_{2V}$}}

\def\HtwoV{\mbox{H$_{2V}$}}

\def\CaIR{\mbox{Ca\,\specchar{ii}\,8542\,\AA}} 

\def\level #1 #2#3#4{$#1 \; ^{#2} \mbox{#3} ^{#4}$}   

\def\degree{\hbox{$^\circ$}}

\def\kms{\hbox{km$\;$s$^{-1}$}}

\def\tis{\!=\!\!}                          

